\documentclass[useAMS,usenatbib]{mn2e}
\usepackage{epsfig,rotate,graphicx}
\usepackage[fleqn]{amsmath}
\usepackage{subfigure}
\usepackage{lscape}
\usepackage{bm}

\newcommand{\mnras}{MNRAS}
\newcommand{\apj}{ApJ}
\newcommand{\aap}{A\&A}
\newcommand{\apjl}{ApJL}
\newcommand{\dd}{\delta}

\newcommand{\be}{\begin{equation}}
\newcommand{\ee}{\end{equation}}
\newcommand{\gtrsim}{\;\raisebox{-.8ex}{$\buildrel{\textstyle>}\over\sim$}\;}
\newcommand{\lesssim}{\; \raisebox{-.8ex}{$\buildrel{\textstyle<}\over\sim$}\;}

\newcommand{\nat}{{\it Nature, }}

\newcommand{\aaps}{{\it A\&AS, }}

\newcommand{\prd}{{\it Phys. Rev. D }}

\title[Migration with unstable gaps]{Outward migration of a giant planet
 with a gravitationally unstable gap edge}

\author[Lin and Papaloizou]{ Min-Kai Lin$^{1,2}$
  \thanks{E-mail: mklin924@cita.utoronto.ca} and John C. B. Papaloizou$^1$ \thanks{E-mail:
    J.C.B.Papaloizou@damtp.cam.ac.uk} \\ 
$^1$ Department of Applied Mathematics and Theoretical Physics,
University of Cambridge, Centre for Mathematical Sciences,\\
\  \ Wilberforce  Road, Cambridge, CB3 0WA, UK \\
$^2$ Canadian Institute for Theoretical Astrophysics,  
60 St. George Street, Toronto, ON, M5S 3H8, Canada \\
}

\begin{document}

\maketitle

\begin{abstract}
We present numerical simulations of disc-planet interactions where the planet opens
a gravitationally unstable gap in an otherwise gravitationally stable disc. 
In our disc models, where the outer gap edge can be unstable 
to global spiral modes, we find that as we increase the surface density scale
the gap becomes more unstable and the planet migrates outwards more rapidly. 
We show that the positive torque is 
 provided by material brought into the planet's coorbital region by
the spiral arms. This material is expected to execute horseshoe turns upon approaching
the planet and hence torque it. Our results suggest that standard type II migration, 
applicable to giant planets in non-self-gravitating viscous discs, is likely to be
significantly modified in massive discs when gravitational instabilities associated
with the gap occur. 
\end{abstract}


\section{Introduction}
One of the common approximations in studies of
disc-planet interactions is to adopt a 
non-self-gravitating disc. Thus, despite over three decades
since the work of \cite{goldreich79,goldreich80}, only a relatively small number of  
studies have adopted self-gravitating disc models 
\citep{nelson03a,nelson03b,pierens05,baruteau08,zhang08,ayliffe10}. 
Disc gravity was included in these works in order to obtain a self-consistent treatment of 
planetary migration. The discs remain gravitationally stable.

Recently, \cite{baruteau11} and \cite{michael11} performed numerical
simulations of planetary migration in gravitationally 
unstable discs. These studies were motivated by the need to understand
the fate of giant planets formed by disc instability \citep{durisen07}. 
Consequently, these authors consider massive discs which are 
gravitationally unstable without the presence of a planet
because the surface density is high enough and the disc is sufficiently cool 
\citep[see][]{toomre64}. 

However, there are other types of 
gravitational instabilities in discs. Of particular relevance to 
disc-planet interactions is gravitational instability associated 
with internal disc edges and grooves \citep{sellwood91,papaloizou91}, as it is known that
giant planets open annular gaps in discs \citep{lin86}. 

\cite{meschiari08} suggested planetary gaps may become gravitationally 
unstable by analysing the stability of a prescribed disc profile. 
Instability was explicitly confirmed by \cite{lin11b} via linear and nonlinear 
calculations for gaps self-consistently opened by a planet. \citeauthor{lin11b}
called these instabilities \emph{edge modes} since they are associated with
gap edges.

In this study we explore the consequence of edge modes
on planetary migration using hydrodynamic simulations. 
We consider the specific configuration of
a giant planet residing in a gap with unstable gap edges.
Our specific aim is to understand how edge modes can 
modify the standard picture of planetary migration expected for  
giant planets.   
We therefore focus on a small 
set of simulations rather than a full parameter survey.


This paper is organised as follows. We describe our
disc-planet models and numerical methods in \S\ref{setup}.  
In \S\ref{model_stability} we discuss the 
expected stability properties of planetary gaps 
formed in our discs. We present migration results in 
\S\ref{overview}. We find that the planet migrates 
\emph{outwards} as the gap edge becomes 
increasingly unstable. We analyse a case in 
\S\ref{Q1d7} to identify the required source of positive torque and 
explain how this can be attributed to edge modes. We conclude
in \S\ref{concl} with a discussion of implications and limitations of 
our results.

\section{Model setup}\label{setup}
We consider a two-dimensional self-gravitating protoplanetary disc of mass $M_d$
orbiting a central star of mass $M_*$. We adopt polar coordinates 
$\bm{r}=(r,\phi)$ centred on the star and a non-rotating reference frame. The
governing hydrodynamic equations are given in \cite{lin11b}. Units are
such that $M_*=G=1$, where $G$ is the gravitational constant. 

The disc occupies $r\in[r_i,r_o]=[1,25]$ and its surface density
$\Sigma$ is initialised to    
\begin{align}   
  \Sigma(r) = \Sigma_0 r^{-3/2}\left[1-\sqrt{\frac{r_i}{r + H(r_i)}}
    \,\right],  
\end{align}
where $H(r)$ is the disc semi-thickness defined below, and
the surface density scale $\Sigma_0$ is chosen by specifying the
Keplerian Toomre $Q$ parameter at the outer boundary
\begin{align}\label{qodef}
  Q_o=\left.\frac{c_\mathrm{iso}\Omega_k}{\pi G \Sigma}\right|_{r_o}, 
\end{align}
where 
\begin{align}
\Omega_k = \sqrt{\frac{GM_*}{r^3}}
\end{align}
is the Keplerian orbital frequency, and 
\begin{align}
c_\mathrm{iso} = H \Omega_k
\end{align}
is the sound-speed profile for a locally isothermal disc. We set $H(r)=hr$ and fix
the aspect-ratio $h=0.05$. 
The disc is also characterised by a uniform
kinematic viscosity $\nu=10^{-5}$ in dimensionless units. The initial
azimuthal velocity $v_\phi$ is set from centrifugal balance with 
stellar gravity, disc gravity and pressure. The initial radial velocity
is set to $v_r = 3\nu/2r$. 

We introduce a planet of fixed mass $M_p = 2\times10^{-3}M_*$ on a
circular orbit at $r=10=r_p(t=0)$. This value of $M_p$
corresponds to 2 Jupiter masses if $M_*=M_\odot$,  which opens a deep
gap leading to type II migration in a typical non-self-gravitating
viscous disc \citep{lin86}, provided no instabilities
develop. In this standard picture, migration follows the viscous
evolution of the gap.

\subsection{Equation of state}
The equation of state (EOS) is locally isothermal, so the vertically
integrated pressure is $p=c_s^2\Sigma$. Before introducing the planet, we set
$c_s = c_\mathrm{iso}$. When a planet is present, the sound-speed $c_s$ is
prescribed as 
\begin{align}\label{pepeos}
  c_s=\frac{hrh_pd_p}{[(hr)^{7/2} + (h_pd_p)^{7/2}]^{2/7}}\sqrt{\Omega_k^2 +
    \Omega_\mathrm{kp}^2}, 
\end{align}
where $\Omega_\mathrm{kp}^2 = GM_p/d_p^3$,  $d_p = \sqrt{
   |\bm{r} - \bm{r}_p|^2+\epsilon_p^2}$ is the softened distance to the
planet, $\bm{r}_p$ being its vector position and $\epsilon_p$ is 
the softening length.   The parameter $h_p$ 
 controls the increase in sound-speed near the planet, relative to the
 $c_\mathrm{iso}$ profile above, and is fixed to $h_p=0.5$.  
 The increase in sound-speed is $c_s/c_\mathrm{iso} = 1.54,\,1.18$ at
  $r_h$ and $2r_h$ away from the planet respectively, where
  $r_h=(M_p/3M_*)^{1/3}r_p$ is the Hill radius.
Far away from the planet, the sound-speed becomes close to 
$c_\mathrm{iso}$. 

This EOS was proposed by \cite{peplinski08a} in a
series of numerical simulations of type III migration in
non-self-gravitating discs. The $c_\mathrm{iso}$ profile, commonly
used in disc-planet simulations, leads to mass accumulation near the
planet and could lead to spurious torques from within the planet's
Hill radius. In a self-gravitating disc,
it also causes the effective planetary mass $M_p'$, which is  $M_p$
plus disc material gravitationally bound to the planet
\citep{crida09}, to increase with the disc surface density scale
\citep{lin11a}. 

Equation \ref{pepeos} increases the
temperature near $\bm{r}_p$, thereby reducing the mass
accumulation and effects above. 
Physically, disc material can be expected to
heat up as it falls into the planet potential and provide a pressure buffer
limiting further mass accumulation. The use of 
equation \ref{pepeos} in the present work is 
a simplified prescription for accounting for this.  However, the gaps
  opened in a disc with $c_s = c_\mathrm{iso}$  
and with equation \ref{pepeos} are similar. The gap widths are
identical and the gap depth is $<5\%$ deeper in the case with $c_s =
c_\mathrm{iso}$.

\subsection{Numerical methods}
The hydrodynamic equations are integrated using the \texttt{FARGO}
code \citep{masset00a,masset00b,baruteau08}. The disc is divided into $N_r\times
N_\phi = 1024\times 2048$ zones in radius and azimuth, spaced logarithmically
and uniformly respectively. The resolution is approximately  16
cells per $H$ (or 28 cells per $r_h$). The cells are nearly square ($\Delta
r/r\Delta\phi = 1.02$). An open boundary condition is applied at $r_i$ and a 
non-reflecting boundary condition at $r_o$ (as used by \cite{zhang08}, see also
\cite{godon96}). Self-gravity is solved via a 2D Poisson integral, in
which a softening length $\epsilon_g = 0.3H$ is set to prevent
divergence and approximately account for the disc's vertical
thickness \citep{baruteau08}. 

The planet is introduced with zero mass at time $t=20P_0$, where 
$P_0= 2\pi/\Omega_k(r_p(t=0))$ is the Keplerian period at the planet's initial
orbital radius. 
Its mass is then increased smoothly over 
$10P_0$ to its final value $M_p$. The planet is then allowed to
respond to disc gravitational forces for $t>30P_0$. A standard fifth
order Runge-Kutta scheme is used to integrate its equation of
motion. The softening length for the planet potential is $\epsilon_p =
0.6H$.

\section{Gap stability}\label{model_stability}
We first consider gap profiles in disc models with 
$Q_0= 1.5,\, 1.7,\, 2.0$ to assess their stability. 
These correspond to Keplerian Toomre values at the planet's initial radius
of $Q_p = 2.77,\, 3.14, \, 3.70 $ and total disc masses of
$M_d/M_*= 0.080,\, 0.071, 0.060$, respectively. The discs are
gravitationally stable to axisymmetric perturbations.  
For smooth radial profiles, they are also expected to be 
gravitationally stable to non-axisymmetric perturbations near $r_p$
because $Q_p > 1.5$. 

The fundamental quantity for stability discussion in a structured barotropic disc is 
the vortensity\footnote{Strictly speaking, our discs are not barotropic. However, the instabilities of interest are
associated with an internal structure with characteristic thickness $ H\ll r $, but 
the sound-speed varies on a global scale and can be approximated as constant, i.e. strictly isothermal and hence
barotropic. } \citep[e.g.][]{papaloizou89,lovelace99}
\begin{align} 
\eta = \frac{\kappa^2}{2\Omega\Sigma}, 
\end{align}
where $\kappa^2 = r^{-3}d(r^4\Omega^2)/dr$ is square of the epicycle frequency and
$\Omega = v_\phi/r$ is the angular velocity. For gaps opened by a giant planet,
the vortensity profile closely follows the Toomre $Q$ profile when the latter is calculated
with $\kappa $ instead of  $\Omega_k$. 
Specifically, local extrema in $\eta$ and $Q$ occur approximately 
at the same radii \citep{lin11b}. 

\cite{lin11b} have shown that 
planetary gaps in massive discs are unstable to global edge modes associated with 
$\mathrm{max}(\eta)$ or  $\mathrm{max}(Q)$. For the adopted disc models, 
the background Keplerian Toomre parameter decreases with $r$, 
i.e. the disc is more self-gravitating at larger radii. 
Thus edge modes are mostly associated with the outer disc ($r>r_p$), as found by 
\cite{lin11b}. 

\begin{figure}
\centering
\includegraphics[scale=0.41,clip=true, trim=0cm 0.84cm 0cm 0cm]{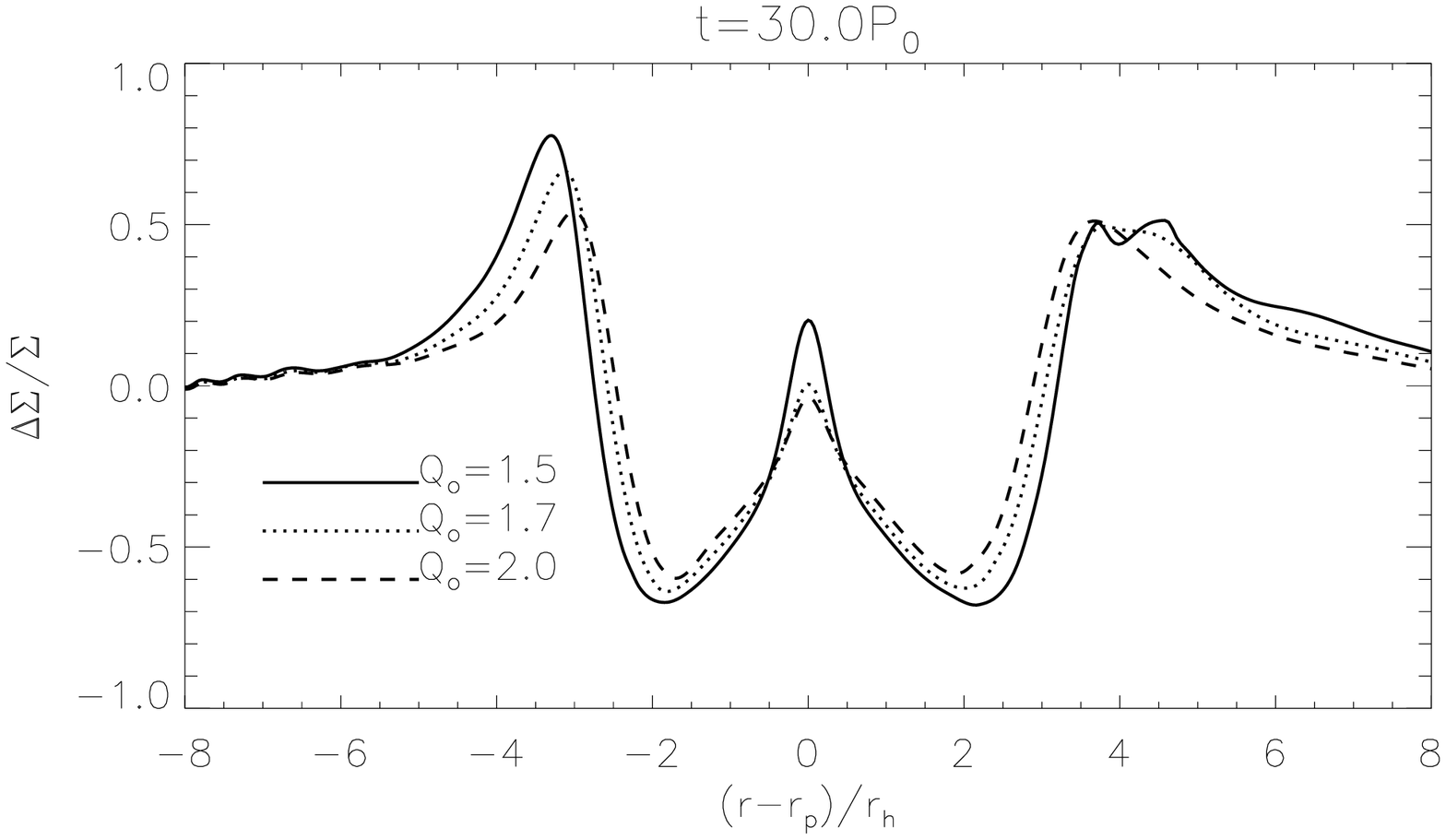}\\
\includegraphics[scale=0.41,clip=true, trim=0cm 0cm 0cm 0.84cm]{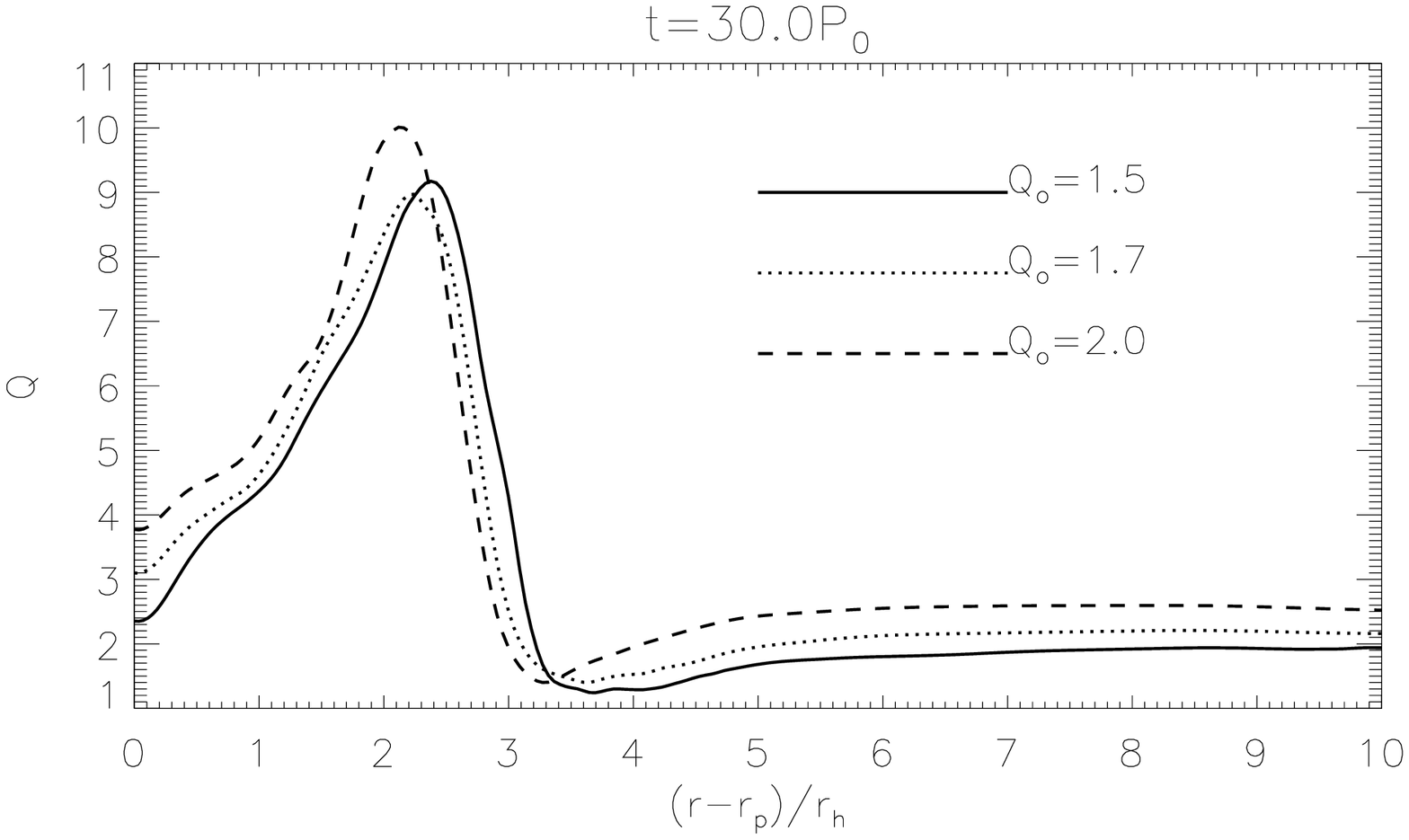}    
\caption{One dimensional profiles of the azimuthally averaged relative surface density perturbation (top)
 and Toomre $Q$ in the outer disc (bottom). Edge modes require
 a sufficiently peaked $\mathrm{max}(Q)$. In our disc models, 
 instability is most prominent in $r>r_p$. 
\label{ch4_1D_profile}}
\end{figure}

Fig. \ref{ch4_1D_profile} compares the  relative surface density
perturbation and the Toomre $Q$ parameter 
for gaps in the above disc models. 
We expect edge modes to have corotation
radius $r_c$ at a vortensity maximum, or about $2r_h$ from the planet. This corresponds
to a local minimum in relative surface density perturbation
 just inside the gap. 
Assuming
the coorbital region of a giant planet is such that $|r - r_p|\lesssim 2.5r_h$ 
\citep{artymowicz04b,paardekooper09}, $r_c$ is just within the outer gap edge. 
This suggests the development of edge modes could
bring over-densities into the coorbital region of the planet. 

Edge modes 
also require coupling to the outer smooth disc via self-gravity, which is easier with
lower $Q_o$. 
The profiles here indicate edge modes will develop more
readily with increasing disc mass and therefore have more significant
effect on migration as $Q_o$ is lowered. 

Differences in gap profiles are also attributable to different effective
planet masses, $M_p'$. Without carefully choosing $M_p$ and the EOS parameter 
$h_p$, it is not possible to have exactly the same $M_p'$ in discs of varying $\Sigma_0$. 
$M_p'$ typically increases with increasing $\Sigma_0$, which 
promotes instability because gap edges become sharper. However, 
increasing surface density generally favours gravitational instability. 
Thus, planetary gaps in discs with decreasing 
$Q_o$ are increasingly unstable, though perhaps more so than if $M_p'$
is held fixed.



 Giant planets can also induce fragmentation in
  self-gravitating discs, a phenomenon previously reported in SPH
  calculations \citep{armitage99,lufkin04}. Indeed, for a test run
  with $Q_o=1.3$, we found the planetary wake, as well as the disc,
  fragments into clumps and no annular gap can be identified. 
  In this case migration is expected to be strongly affected by
  clumps, instead of large-scale spiral arms which we will focus on
  below.

\section{Migration and disc structure}\label{overview}
The above discussion suggests that edge modes could lead to
 non-axisymmetric disturbances inside the gap.  
Torques may then originate from within the
planet's coorbital region. Because $r_c$ is expected near 
the outer edge of the coorbital region, edge mode over-densities 
may undergo horseshoe turns upon approaching the planet.   
Furthermore, fluid elements just inside the separatrix 
will traverse close to the planet when
executing U-turns, and this may provide significant torques. 
With this in mind, in this section we aim to identify the correlation between
migration and gap evolution. 

\begin{figure}
  \centering
  \includegraphics[width=\linewidth]{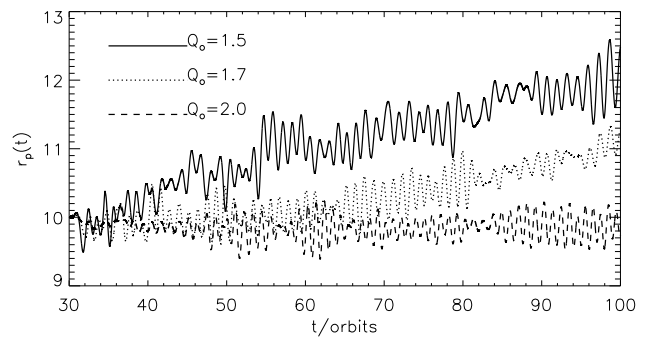}
  \caption{Migration of a giant planet in massive discs.
    The $Q_o=1.5$ and $Q_o=1.7$ models develop edge modes throughout, whereas the
    $Q_o=2.0$ model first develops vortex modes.
    \label{ch4_migration}}
\end{figure}

\begin{figure}
\centering
\includegraphics[width=\linewidth]{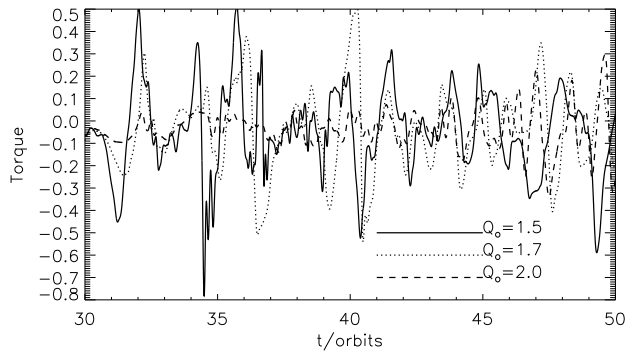}
\caption{Instantaneous disc-on-planet torques. We have made this plot
 more comprehensible by multiplying the torques by a factor
$f=1-\exp{\left(-|\bm{r}-\bm{r_p}|^2/r_h^2\right)}$. This reduces local contributions
from the Hill sphere but does not affect the important feature --- rapid oscillatory torques --- or
their behaviour as a function of $Q_o$. Despite tapering, instabilities still have significant impact on
disc-planet torques.
\label{ch4_torque_instant}}
\end{figure}

Fig. \ref{ch4_migration} shows the instantaneous orbital radius of the
planet in the above disc models. Migration is non-monotonic
and can be inwards or outwards on short timescales ($\sim P_0$). However,
with increasing disc mass, outward migration is favoured on 
timescales of a few 10's of orbits. For $Q_o=1.5$, $r_p$ increases by $20\%$
in only $70P_0$. This is distinct from standard type II migration,
which is inwards and occurs on much longer, viscous timescales 
($t_\nu = r^2/\nu =O(10^4P_0)$ for our choice of $\nu$).

Fig. \ref{ch4_torque_instant} shows the  instantaneous disc-on-planet
torques during the first $20P_0$ after releasing the planet. Instabilities
develop within this time frame and cause the torque to be 
positive or negative at a given instant. 
Up to about $t=40P_0$,  
the torques for $Q_o=1.5$ and $Q_o=1.7$ both show large and rapid oscillations 
in comparison to the $Q_o=2.0$ case. We shall see that this is related to edge modes developing 
for $Q_o=1.5,\,1.7$ but not for $Q_o=2.0$.

\subsection{Disruption of the outer gap edge}

\begin{figure*}
\centering
\includegraphics[scale=.6,clip=true,trim=0cm 0.0cm 1.65cm
0cm]{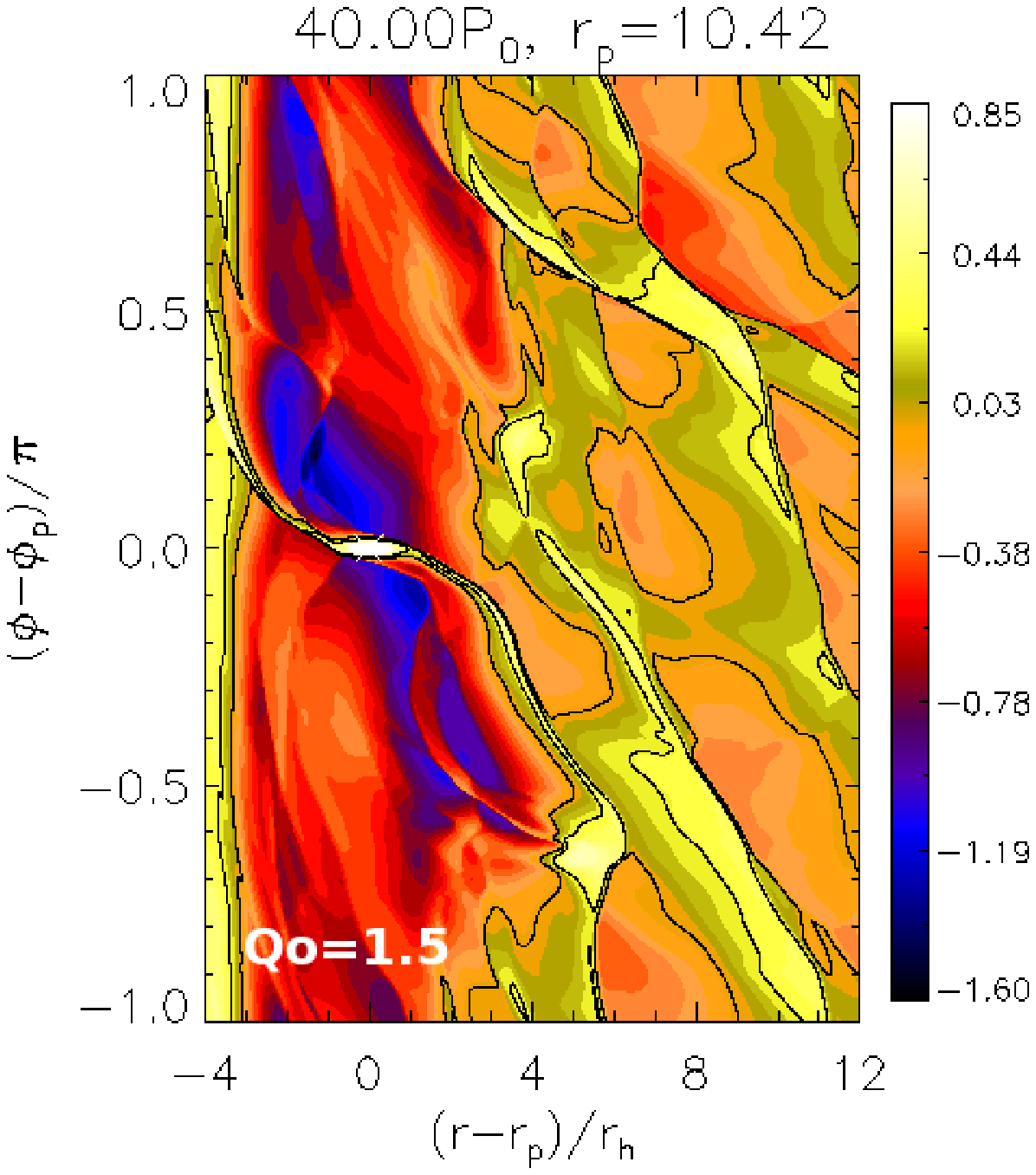}\includegraphics[scale=.6,clip=true,trim=2.13cm
0cm 1.65cm
0cm]{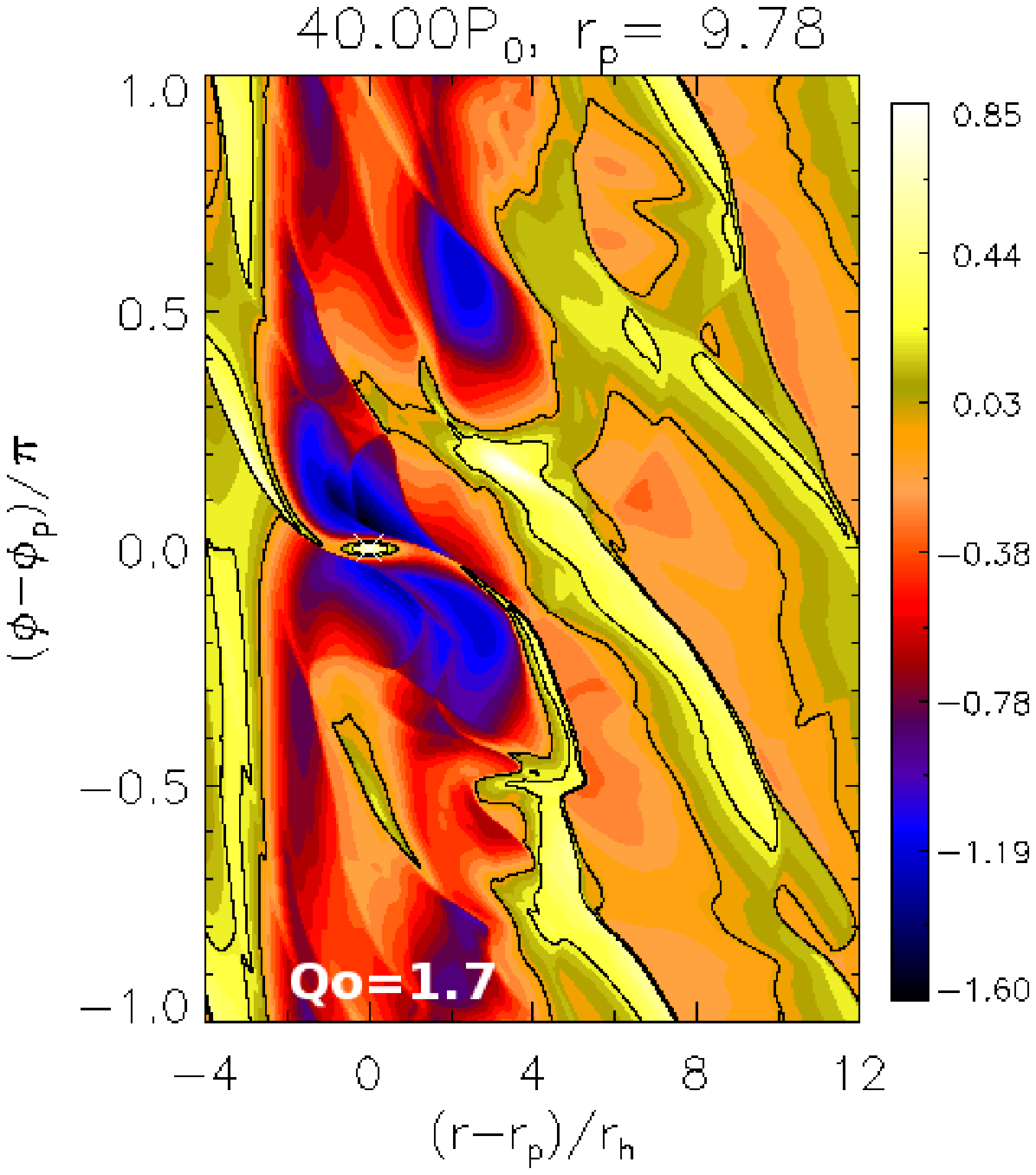}\includegraphics[scale=.6,clip=true,trim=2.13cm
0cm 0cm 0cm]{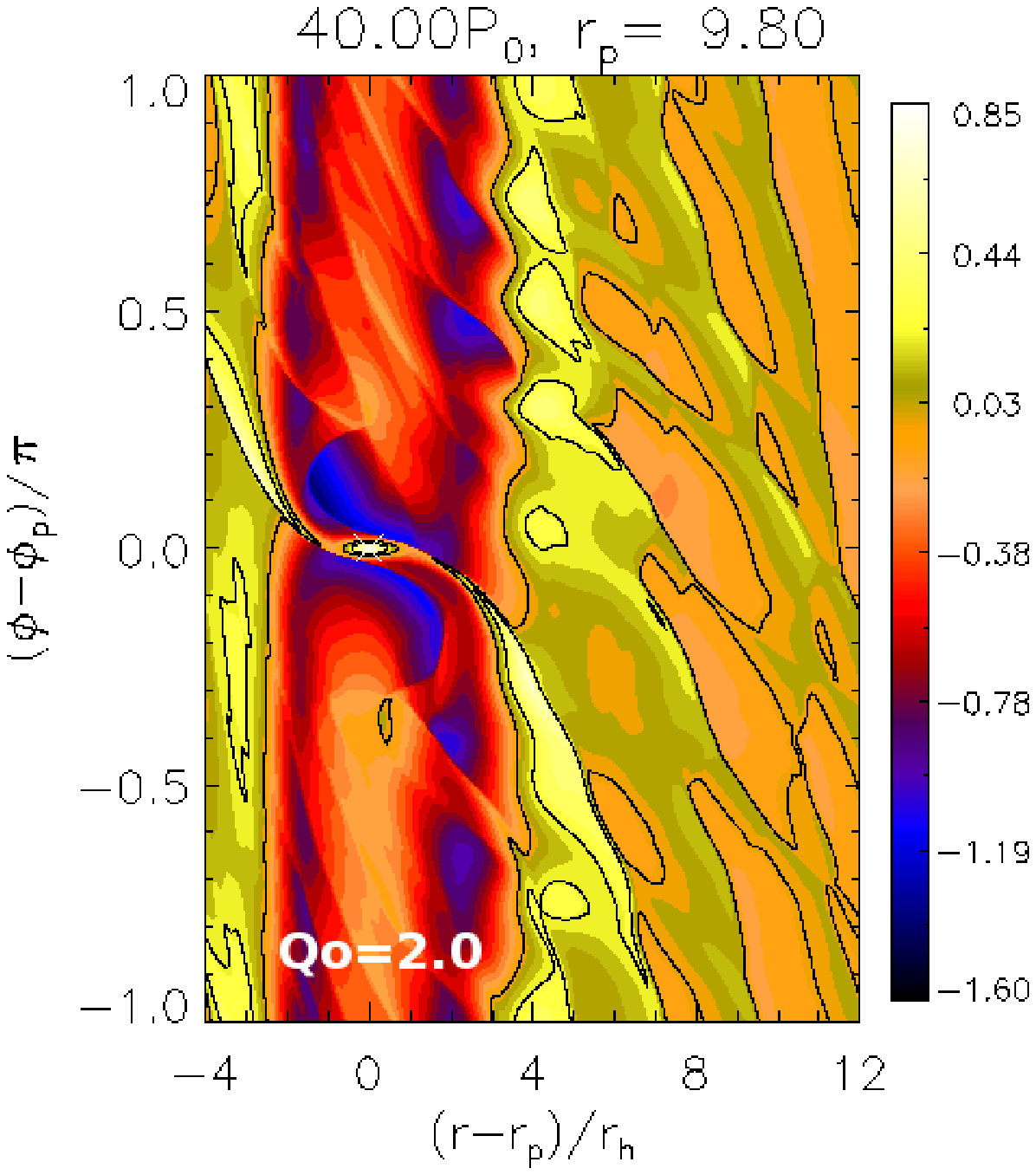}
\caption{Logarithmic relative surface density perturbation,
   $\log{\left[\Sigma/\Sigma(t=0)\right]}$, showing gap
   instability as a function of $Q_o$. As $Q_o$ is lowered, the outer gap
edge becomes more unstable. This correlates with an increasing tendency for outward migration. 
\label{ch4_polarxy_t40}} 
\end{figure*}

The behaviour of $r_p(t)$ correlates with gap structure,  
particularly the outer gap edge.   
Fig. \ref{ch4_polarxy_t40} shows the relative surface density
perturbation when instabilities first develop. The least massive disc with 
$Q_o=2.0$ develops vortices rather than global edge modes (cf. $Q_o=1.7$). 
This can be expected since edge modes require sufficiently strong self-gravity 
\citep{lin11b}.

In weakly or non-self-gravitating structured discs, 
unstable modes are associated with local vortensity minima \citep{lin11a} 
and lead to vortex formation. The basic state (Fig. \ref{ch4_1D_profile})
shows that local $\mathrm{min}(Q)$, hence $\mathrm{min}(\eta),$ 
is located exterior to $\mathrm{max}(Q)$, and is most pronounced for
$Q_o=2.0$. We did observe spiral arms to develop in $Q_o=2.0$ later on, 
but these probably resulted from the vortices perturbing the disc, rather than
the linear edge mode instability. The important point with $Q_o=2.0$ is that
instability leaves the outer gap edge intact and identifiable, 
unlike for more massive discs.

In Fig. \ref{ch4_polarxy_t40}, the $Q_o=1.7$ case develops a $m=2$---3 
edge mode. Protrusion of the spiral arms into the gap makes 
the outer gap edge less well-defined. The surface density in the
  gap is on average higher in $\phi > \phi_p$ than in $\phi < \phi_p$.  
  The edge mode spiral 
  inside the gap and just upstream of the planet could provide
  a significant positive torque as the spiral pattern approaches the
  planet (from above in the figure). This is consistent with  
the large positive torque around the time of the chosen snapshot seen in 
Fig. \ref{ch4_torque_instant}.  
The outer gap edge for $Q_o=2.0$ is not as disrupted 
and this results in no secular increase in $r_p$ for the simulated 
time for $Q_o=2.0$.

In Fig. \ref{ch4_polarxy_t40}, large-scale spirals can still be seen
for $Q_o=1.5$. Weak fragmentation occurs without a collapse into
clumps. Like $Q_o=1.7$, 
the gap is unclean with significant disruption to the outer gap
edge. It is no longer a clear feature as for $Q_o=2.0$. 

The surface density plots (Fig. \ref{ch4_polarxy_t40}), 
together with the torque plot (Fig. \ref{ch4_torque_instant}), 
show that edge modes significantly modify torques from those expected for 
standard type II migration. Specifically for our models, edge modes are associated
with the outer gap edge and lead to a secular increase in orbital 
radius (Fig. \ref{ch4_migration}). We shall see below that this is because
edge modes provide over-densities inside the gap, despite the tendency for
giant planets to clear the gap of material.

\subsection{Gap evolution}\label{gap_evolution}
In this section we use the following 
prescription to examine the evolution of the gap structure. We first calculate
the azimuthally averaged relative surface density 
perturbation,
\begin{align}
\dd\Sigma(r) = \left\langle \frac{\Sigma - \Sigma(t=0)}{\Sigma(t=0)}\right\rangle_\phi. 
\end{align}

\begin{figure}
\centering
\includegraphics[scale=0.42,clip=true,trim=0cm 1.8cm 0cm
0.74cm]{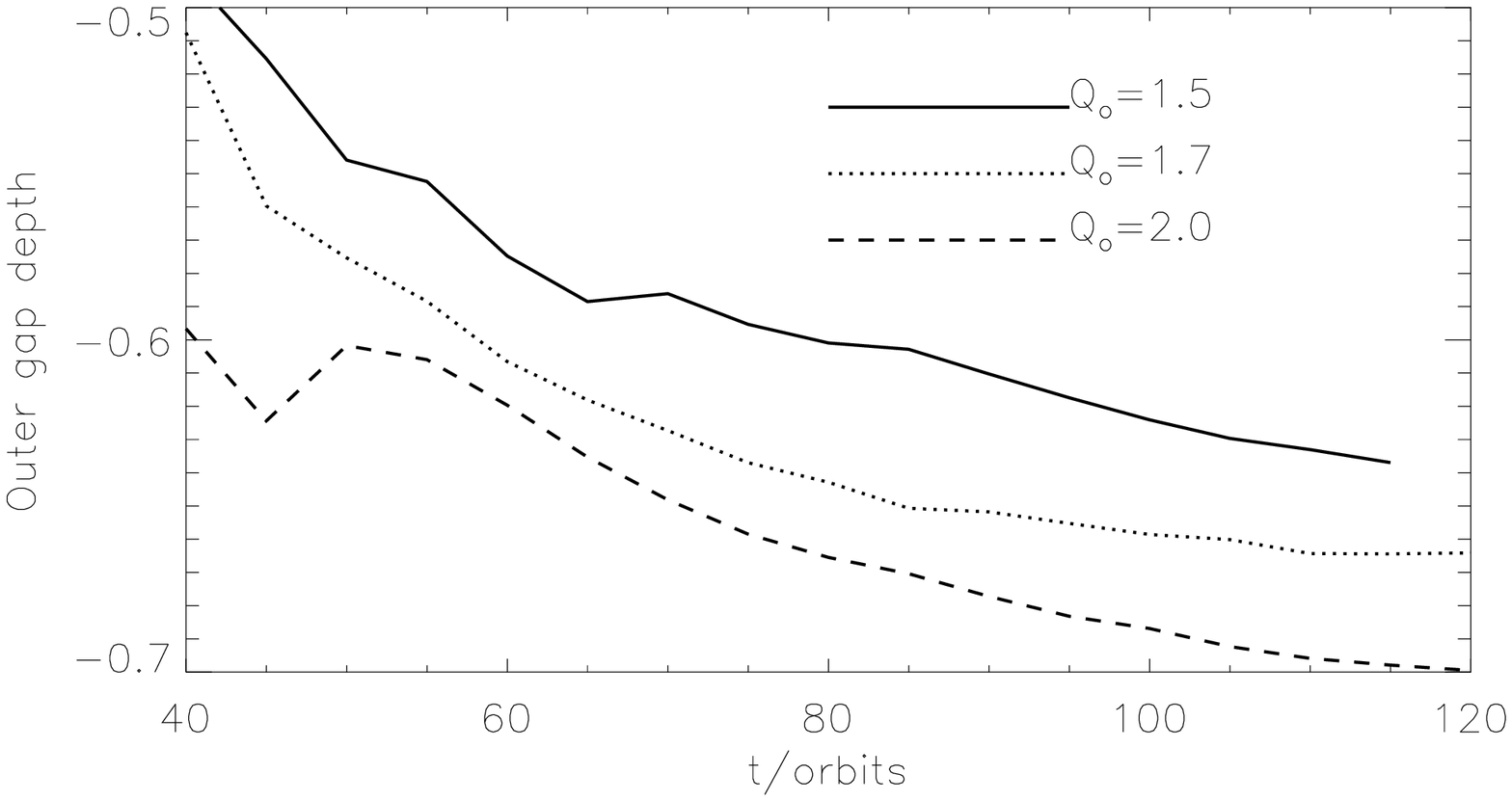}\\
\includegraphics[scale=0.42,clip=true,trim=0cm 0.3cm 0cm 0.74cm]{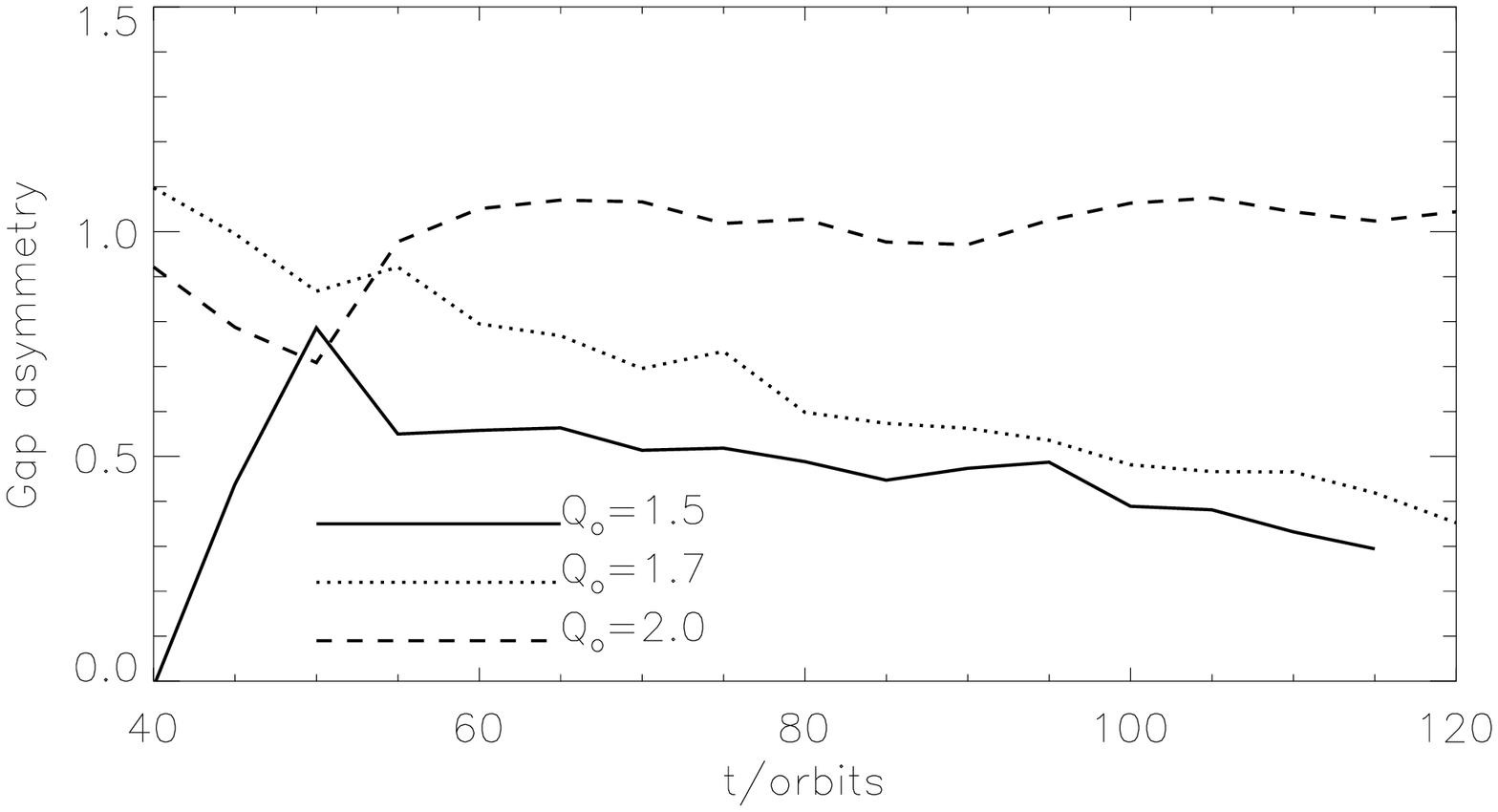}
\caption{Running-time averages of gap properties: the dimensionless
  outer gap depth  (top) and the gap asymmetry in units of Hill
  radius (bottom). These quantities are defined in
  \S\ref{gap_evolution}.
\label{ch4_gap_evolution}}
\end{figure}

We define the \emph{outer gap
  edge} as $r_{e,\mathrm{out}}>r_p$ such that $\dd\Sigma(r_{e,\mathrm{out}}) = 0$,    
and similarly for the \emph{inner gap edge} $r_{e,\mathrm{in}}<r_p$. 
The \emph{outer gap depth} is defined as $\dd\Sigma$ averaged over
$r\in[r_p,r_{e,\mathrm{out}}]$. 
We also define the \emph{outer gap width}, in units of
the Hill radius, as $w_\mathrm{out} = |r_{e,\mathrm{out}} - r_p|/r_h
$ and the \emph{inner gap width} as $w_\mathrm{in} = 
|r_{e,\mathrm{in}} - r_p|/r_h$. The \emph{gap asymmetry} is
$w_\mathrm{out} - w_\mathrm{in}$. 
Running time-averaged plots of these quantities 
are shown in Fig. \ref{ch4_gap_evolution}.

Edge modes have a gap-filling effect. The case $Q_o=1.5$, for which the
planet immediately migrates outwards, has the smallest magnitude of gap depth
throughout. The $Q_o=2.0$ case has the deepest gap and is non-migrating
over the simulation timescale. In the $Q_o=1.7$ case, the magnitude
of gap depth decreases relative to that for 
$Q_o=2.0$ after about $t=80P_0$. Note that this corresponds to outward migration for $Q_o=1.7$.
These observations suggest that material brought into the gap by edge modes, 
is responsible for outward migration.  

We have defined gap asymmetry as the distance
from the planet to the outer gap edge minus the distance between the planet
and the inner gap edge. 
It follows that the more negative the gap asymmetry is,
the closer the planet is to the outer gap edge. 
Fig. \ref{ch4_gap_evolution} shows the planet is located 
closer to the outer gap edge in discs with lower $Q_o$ than in discs with higher $Q_o$. 
Furthermore, the non-migrating $Q_o=2.0$ case reaches a constant asymmetry, 
whereas in the outwards-migrating cases asymmetry decreases with time. 

Recall that edge mode spiral arms are associated with the outer gap edge.  
The trend above suggests that the planet is moving
outer gap edge material inwards via horseshoe turns  
(this would also be consistent with
shallower gaps with decreasing $Q_o$). This provides a
positive torque on the planet.   
If on average this effect dominates over sources 
of negative torques, e.g. Lindblad torques or coincidence of
an edge mode spiral arm with the outer planetary wake \citep{lin11b}, then 
the planet should on average migrate outwards, i.e. closer to the outer gap edge. 




\section{A fiducial case}\label{Q1d7}
\begin{figure}
\centering
\includegraphics[width=\linewidth,clip=true,trim=0cm 0cm 0cm 0.4cm]{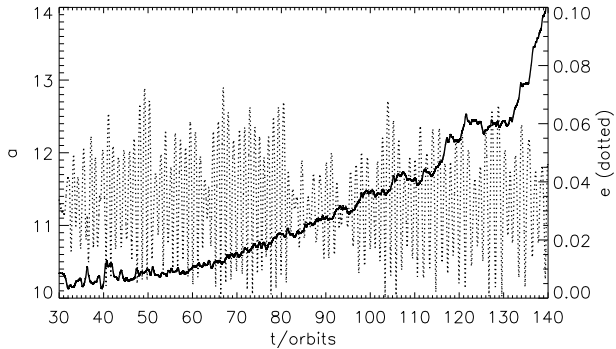}
\caption{Orbital evolution of the planet in the $Q_o=1.7$ disc, in
  terms of Keplerian semi-major axis $a$ (solid) and eccentricity $e$
  (dotted). These have been calculated assuming Keplerian ellipses without accounting for the disc
  potential.
\label{ch4_Q1d7_migration}}
\end{figure}

\begin{figure}
\centering
\includegraphics[scale=.42,clip=true,trim=0cm 1.8cm 1.83cm
0cm]{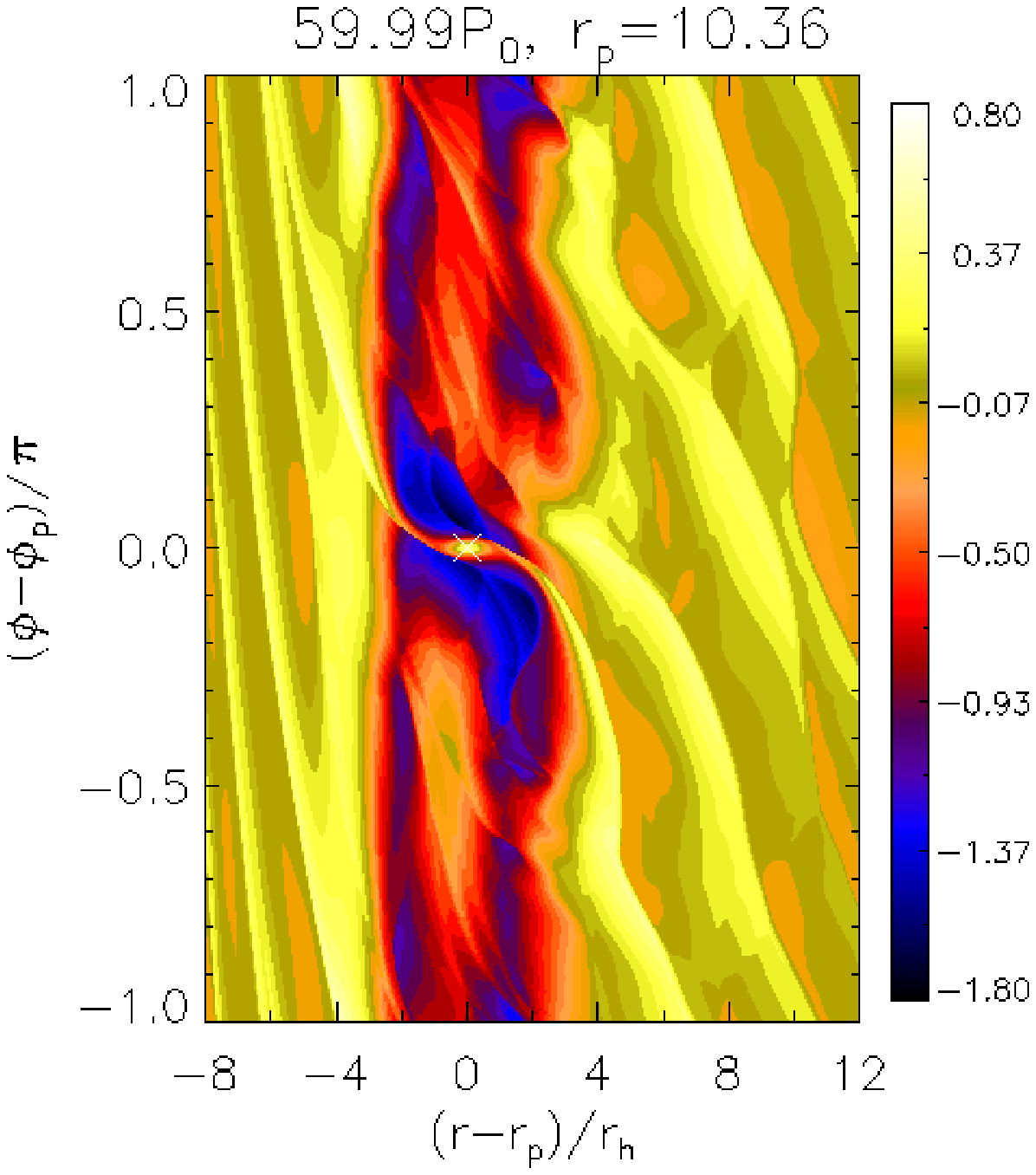}\includegraphics[scale=.42,clip=true,trim=2.23cm
1.8cm 0cm
0cm]{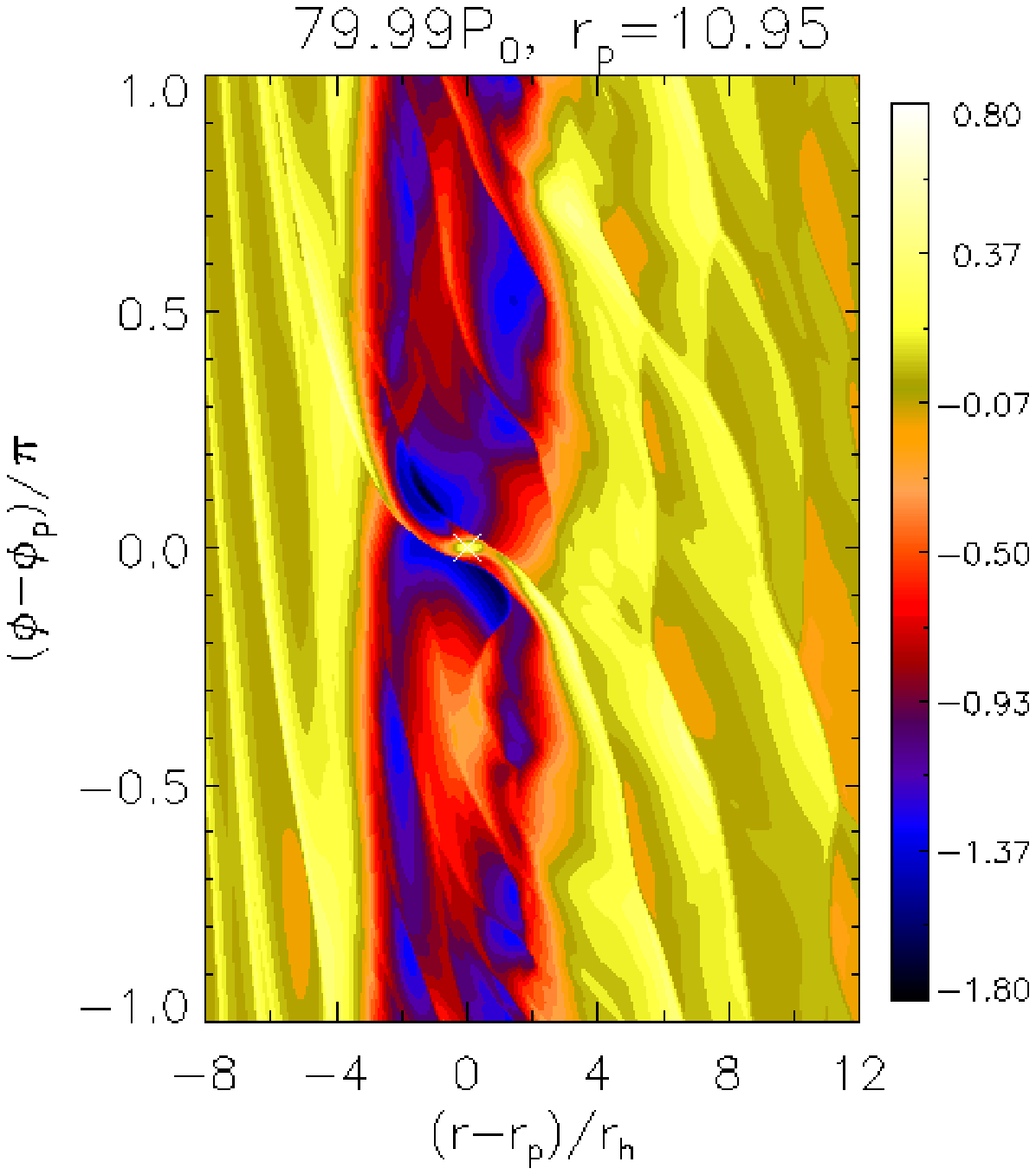}\\
\includegraphics[scale=.42,clip=true,trim=0cm
0cm 1.83cm 0cm]{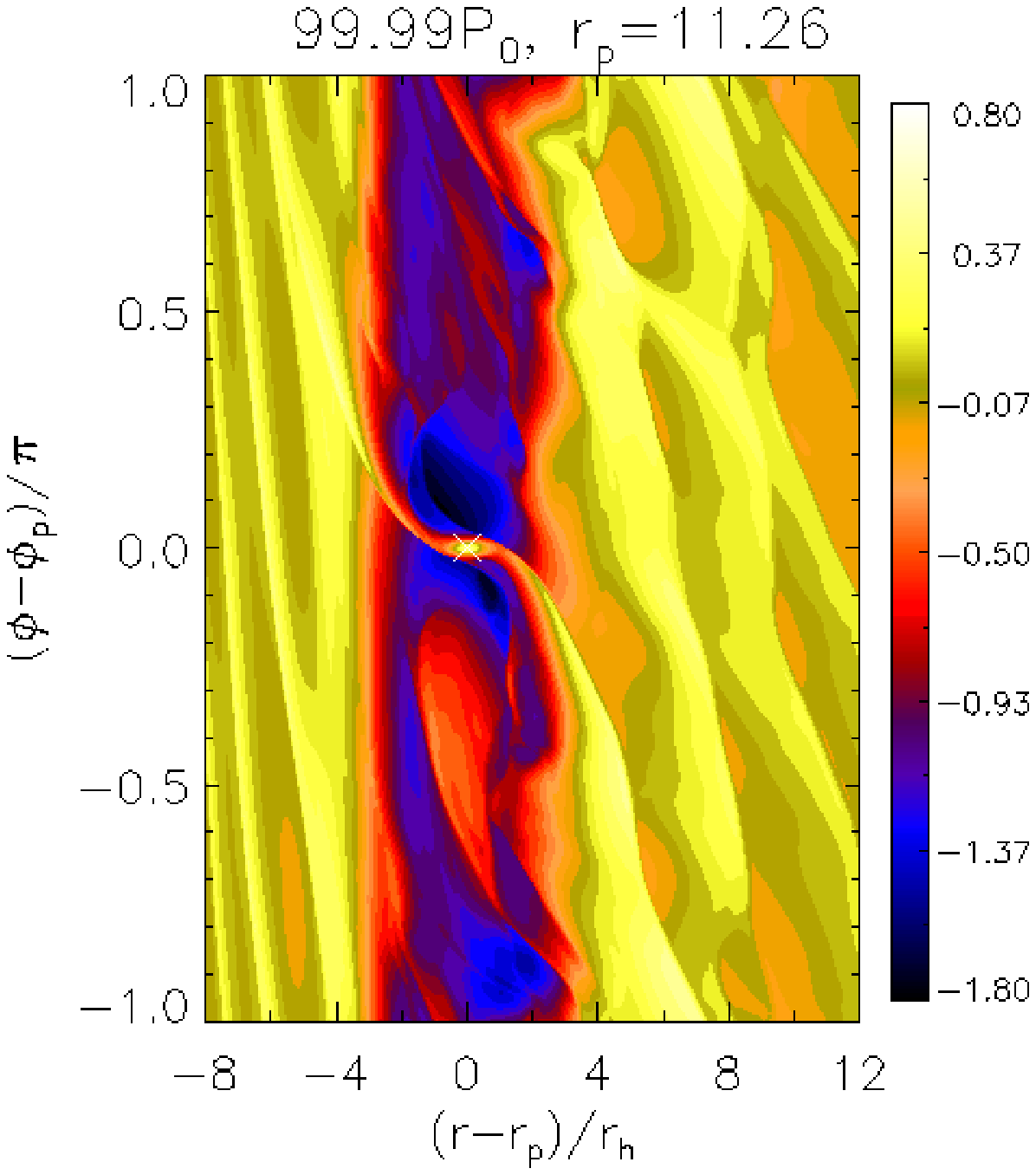}\includegraphics[scale=.42,clip=true,trim=2.23cm
0cm 0cm 0cm]{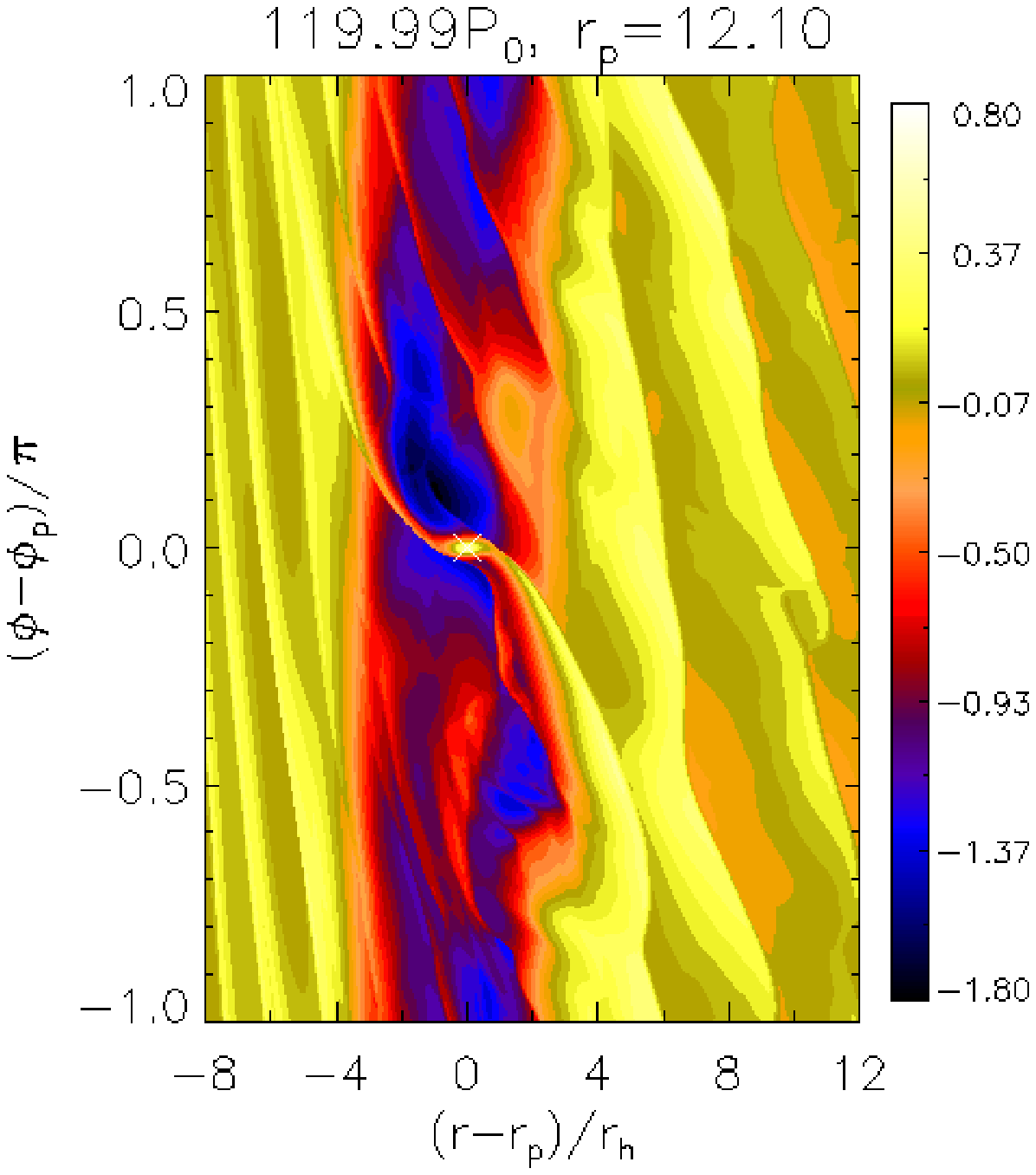}
\caption{Overall evolution of the $Q_o=1.7$ case. The
  logarithmic relative surface density perturbation
   $\log{\left[\Sigma/\Sigma(t=0)\right]}$ is shown.
\label{ch4_Q1d7_polarxy}}
\end{figure}

In order to understand the mechanism by which edge mode
spiral arms lead to outward migration, we here focus on the case
$Q_o=1.7$. This simulation has been extended to $t=140P_0$. 
Fig. \ref{ch4_Q1d7_migration} and Fig. \ref{ch4_Q1d7_polarxy}
summarises this simulation  
in terms of the evolution of the planet's orbit 
and the disc's surface density perturbation. 
The orbit remains fairly circular $(e< 0.06)$ with no
overall change in eccentricity. However, the semi-major axis $a$ increases
more noticeably than the orbital radius $r_p$.

In Fig. \ref{ch4_Q1d7_migration}, $a$ rapidly increases towards
the end of the simulation ($t>130P_0$). This is because edge
  modes eventually 
  caused the planet to move into the disc lying 
beyond the original outer gap edge. The planet effectively leaves the
gap. At this point, we found the surface density contrast ahead and behind the planet
conforms to outward type III migration
\citep{masset03,peplinski08c}. This may have resulted from the fact
that edge modes in our discs supply
over-densities ahead of the planet (see below),  providing the
initial condition, or kick, for outward runaway
migration.

Once the planet is in type III migration, edge modes become irrelevant. 
This situation differs from migration sustained by edge mode spirals 
associated with the gap edge, where the planet can still be seen to reside in an
annular gap. Henceforth we shall focus on the time frame in which this
configuration holds ($t\lesssim 130P_0$).

Fig. \ref{ch4_Q1d7_polarxy} shows the outer disc ($r>r_p$) is
more unstable than the inner disc ($r<r_p$). 
Large-scale, coherent spirals are maintained throughout 
the simulation. The gap is unclean with material
brought into it by the edge modes. 
Over-densities may lie within the planet's coorbital region ($2.5r_h\gtrsim r-r_p > 0$). 
Such material is expected to execute inward horseshoe turns and exert a positive torque
on the planet. 

 The above effect causes the planet to move outwards,  
towards the outer gap edge where the surface density is higher. The 
planet can then interact with that material by moving it
through inward horseshoe turns, providing further positive
torque. The torque magnitude should also increase with the migration
speed. These effects can result in a positive feedback
akin to that seen in classic type III migration \citep{masset03},
which is consistent with the faster-than-linear increase in $a$ for 
$t<130P_0$. 

This interaction is local and depends
primarily on the surface density, so the background Toomre $Q$
parameter, which decreases with radius on a global scale, is not
expected to be of direct relevance in this feedback mechanism apart
from its dependence  on the surface density. 


\begin{figure}
\centering
\includegraphics[scale=.24,clip=true,trim=0cm 1.8cm 1.83cm
0cm]{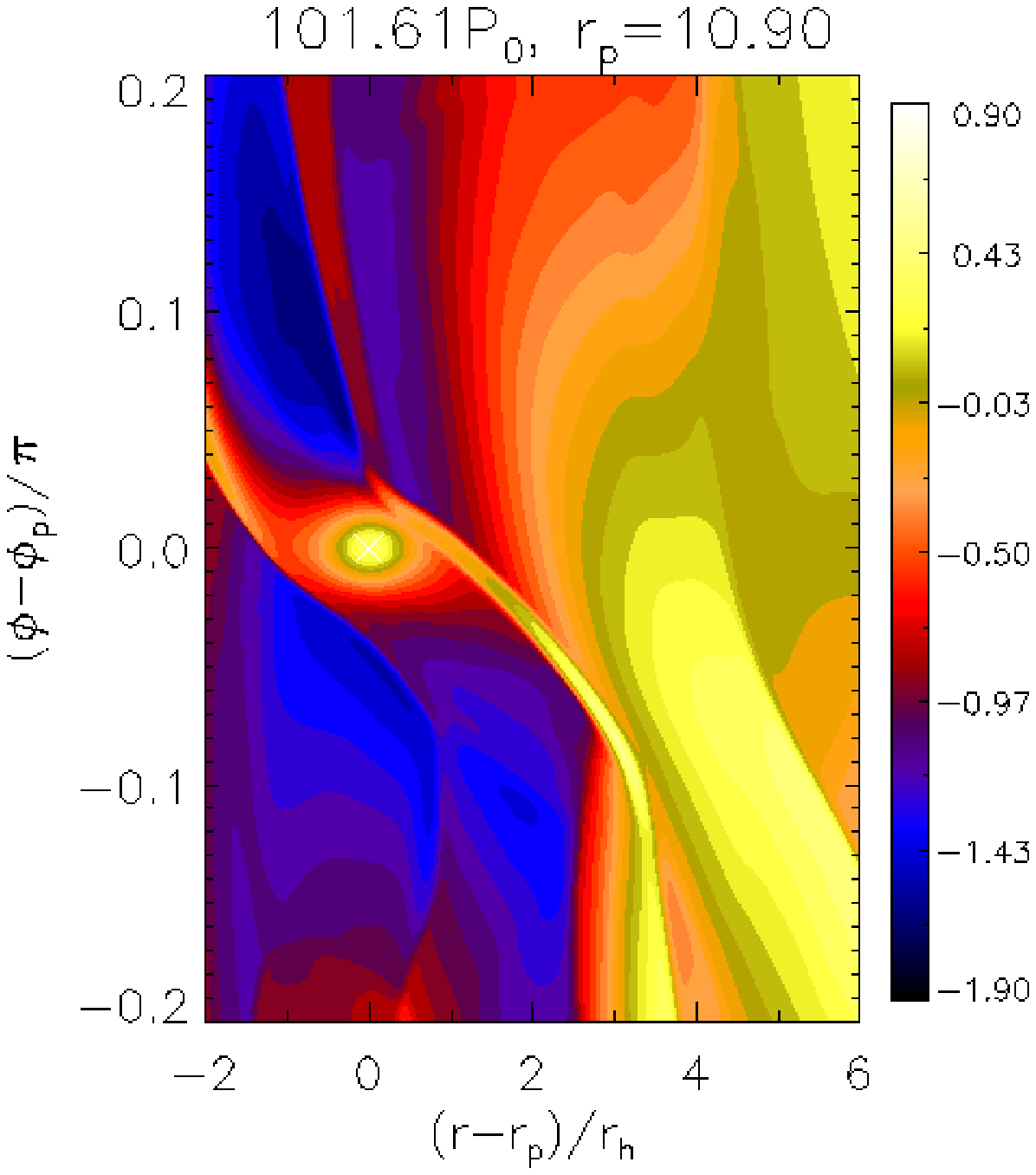}\includegraphics[scale=.24,clip=true,trim=2.23cm
1.8cm 1.83cm
0cm]{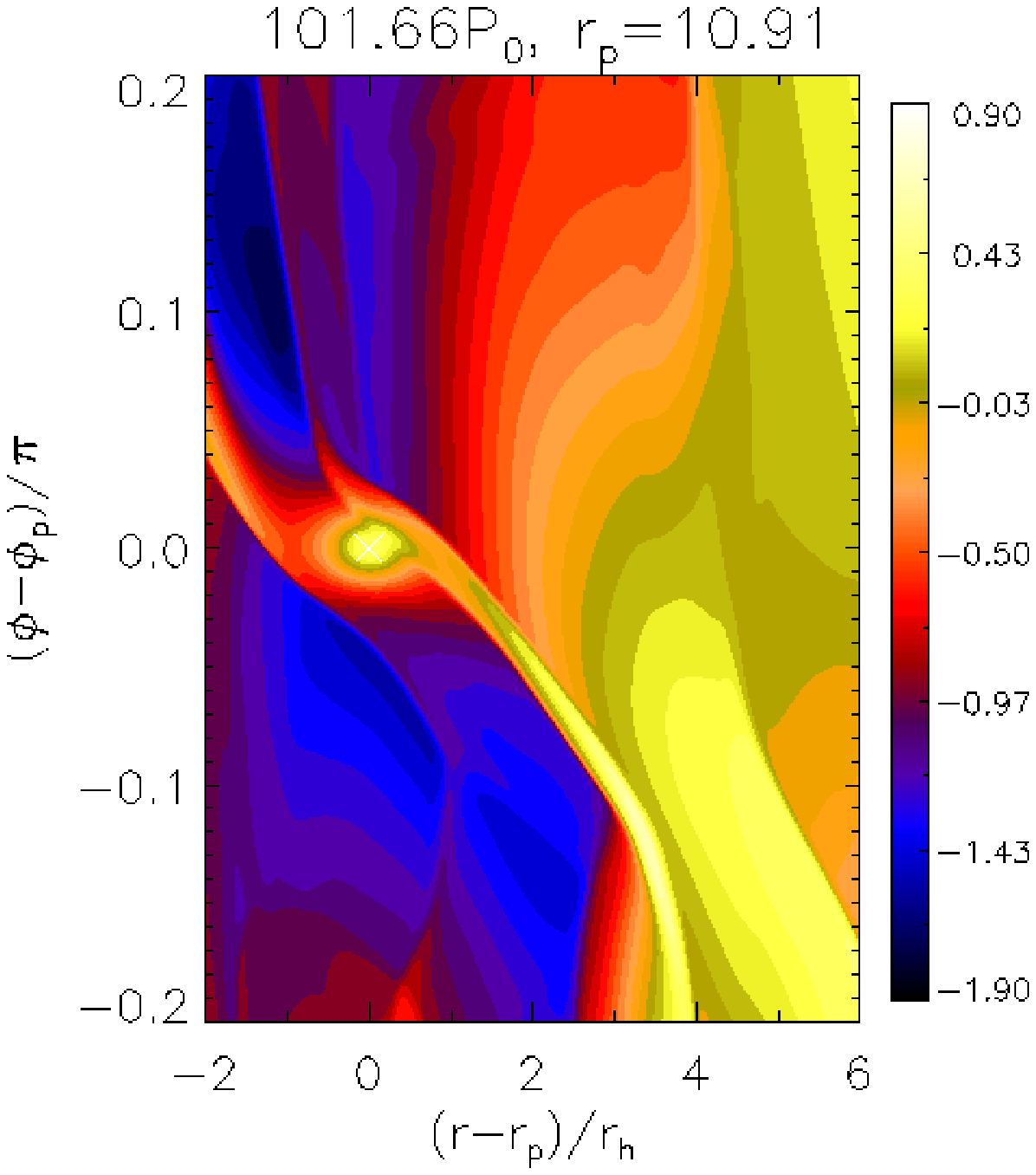}\includegraphics[scale=.24,clip=true,trim=2.23cm
1.8cm 1.83cm 0cm]{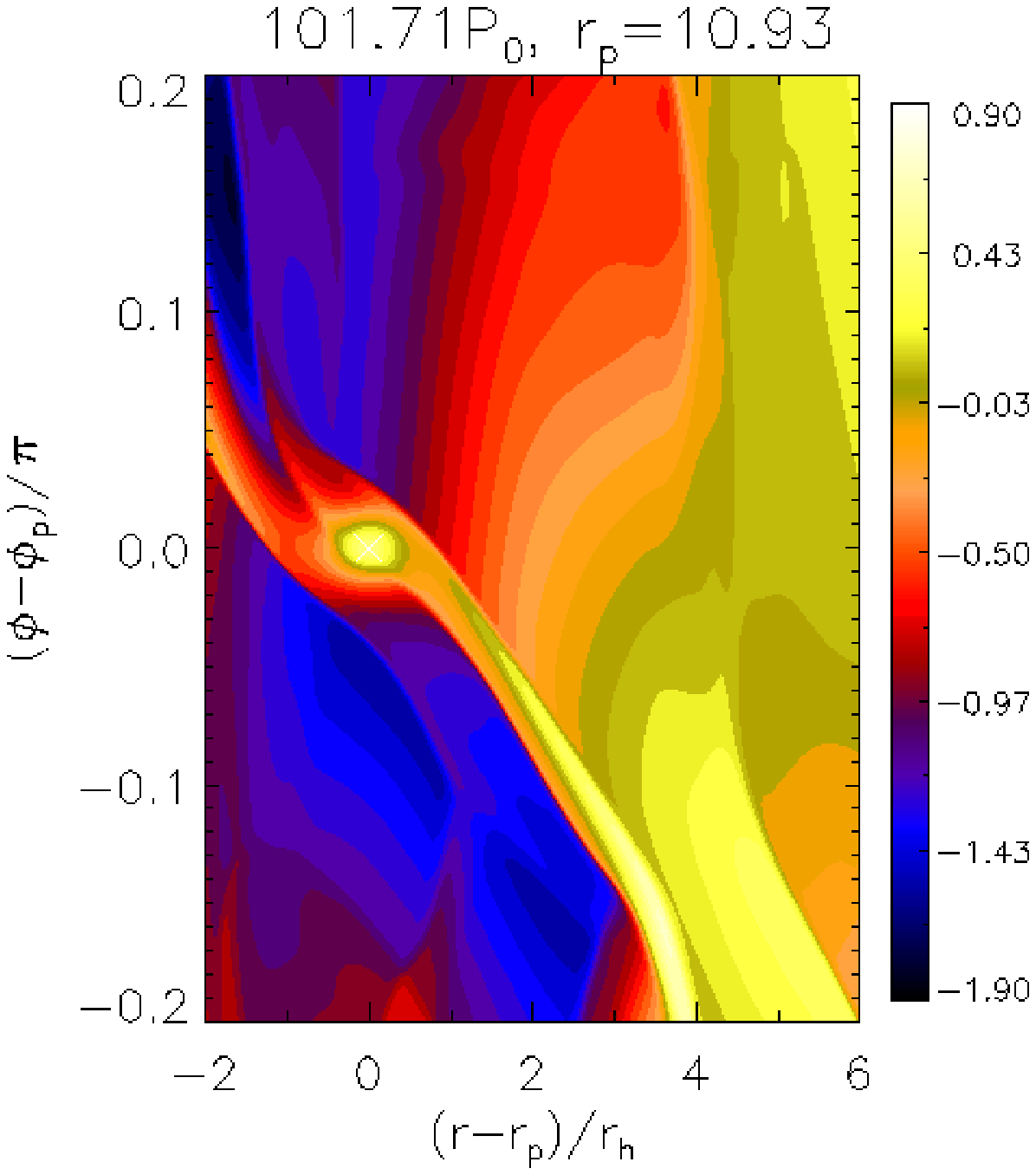}\includegraphics[scale=.24,clip=true,trim=2.23cm
1.8cm 0cm 0cm]{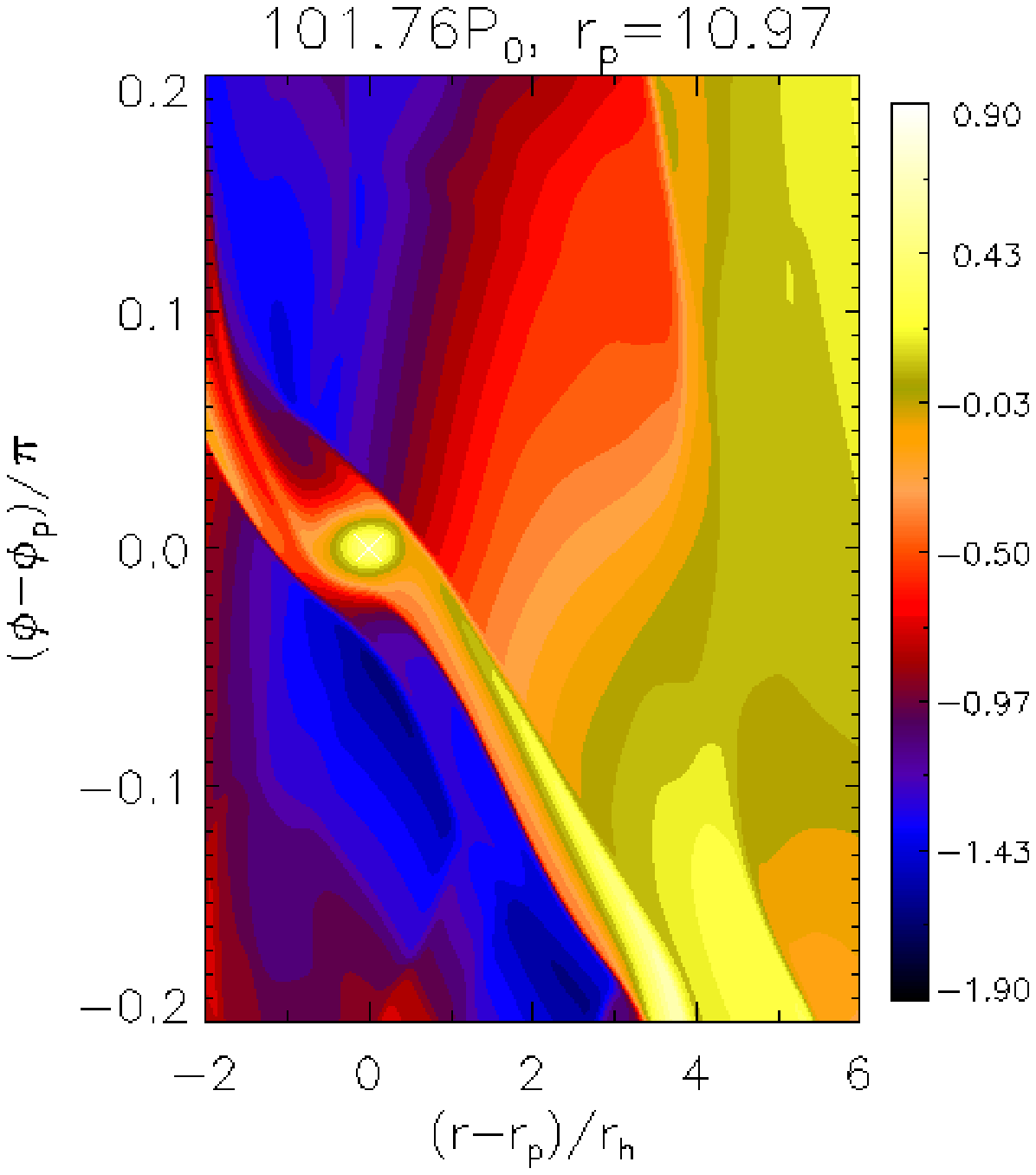}\\
\includegraphics[scale=.24,clip=true,trim=0cm 0cm 1.83cm
0cm]{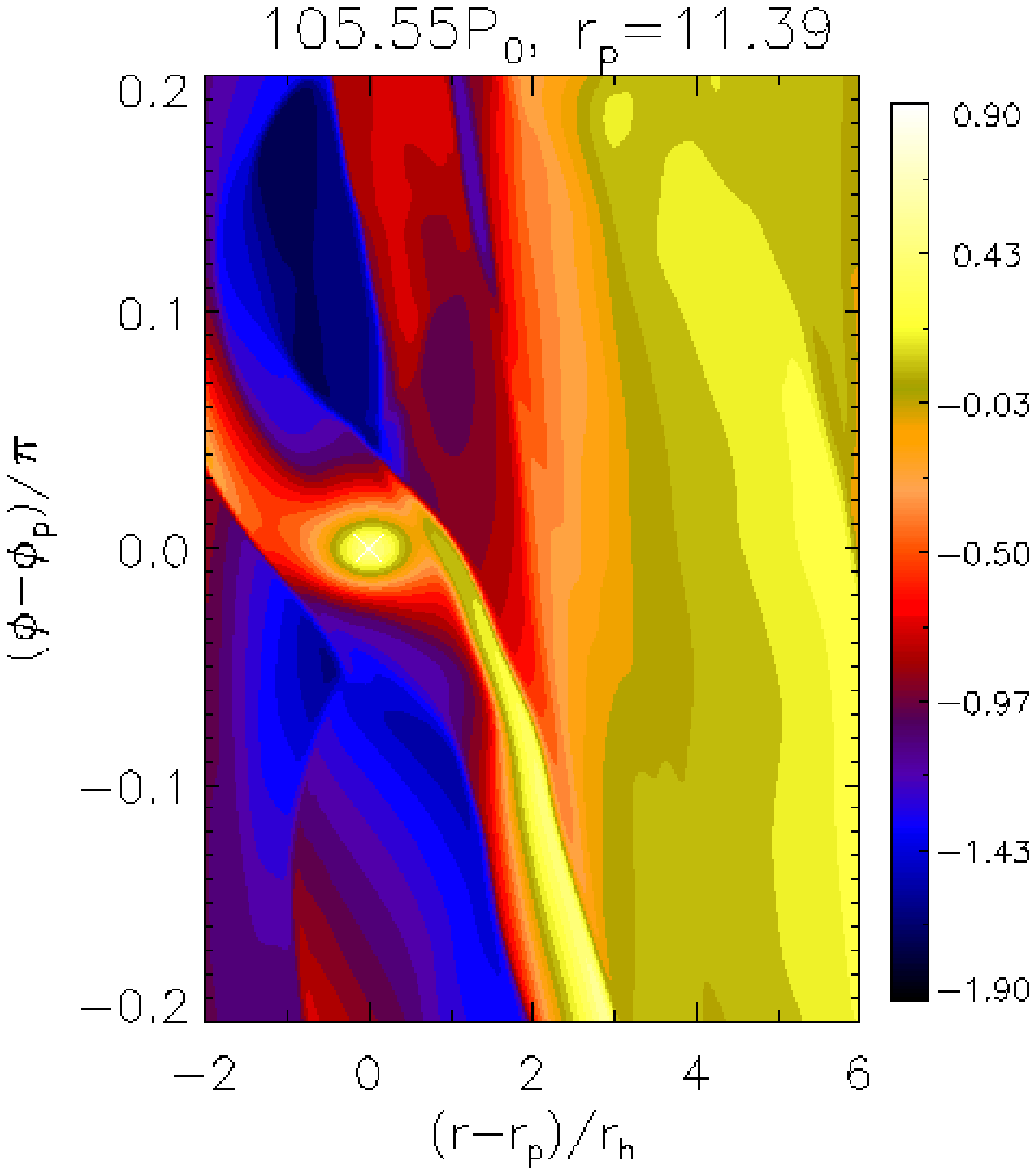}\includegraphics[scale=.24,clip=true,trim=2.23cm
0cm 1.83cm
0cm]{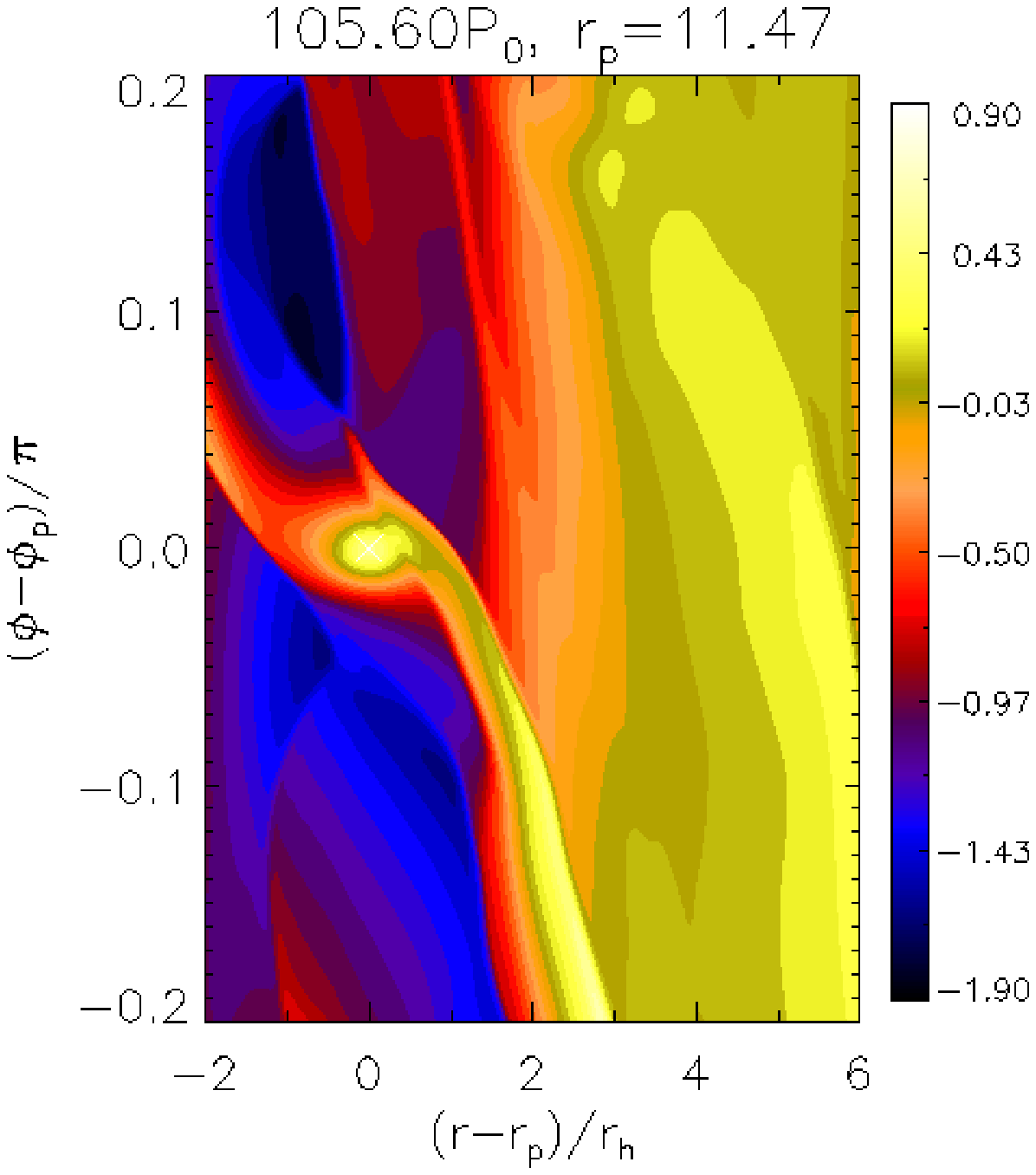}\includegraphics[scale=.24,clip=true,trim=2.23cm
0cm 1.83cm 0cm]{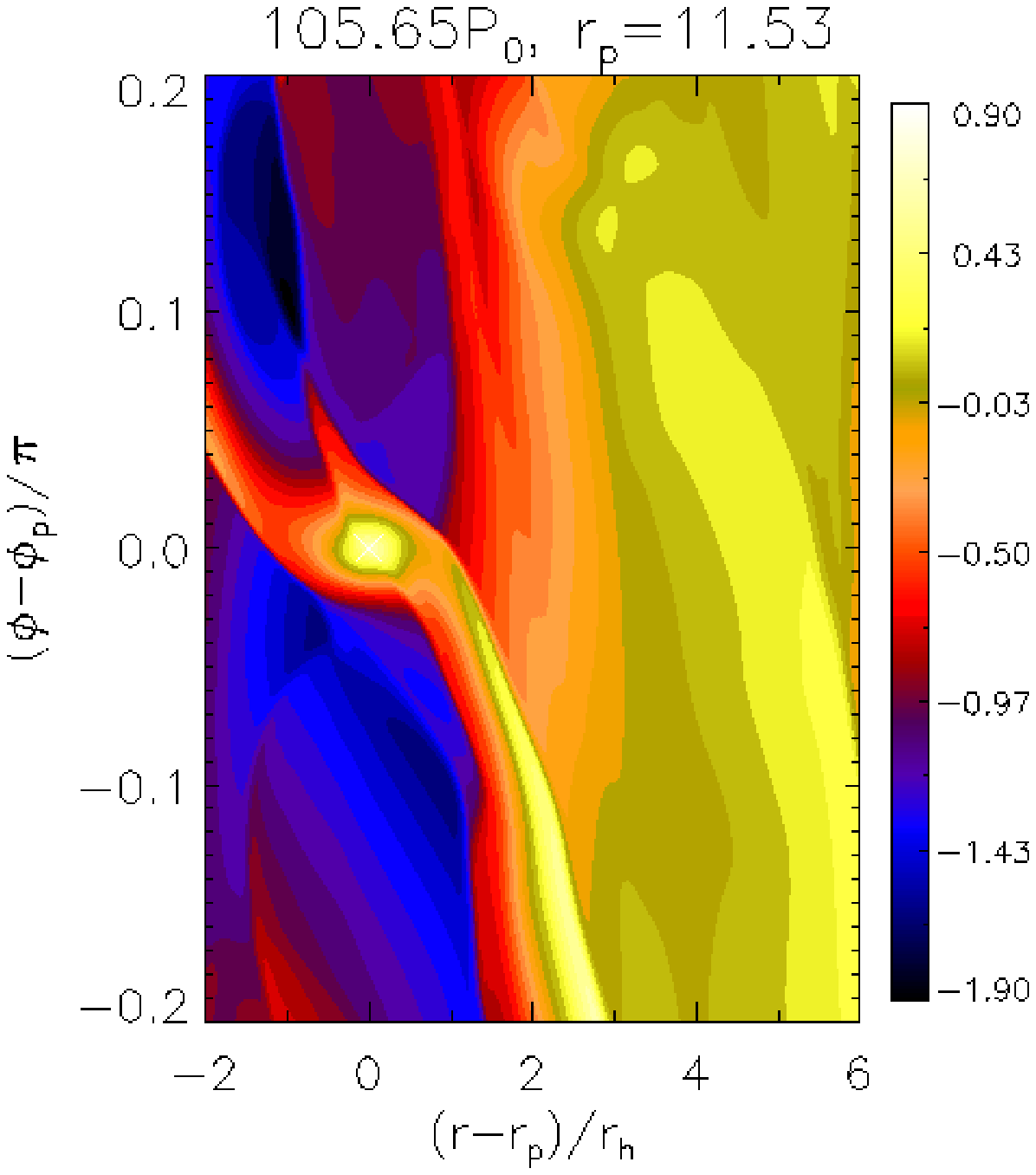}\includegraphics[scale=.24,clip=true,trim=2.23cm
0cm 0cm 0cm]{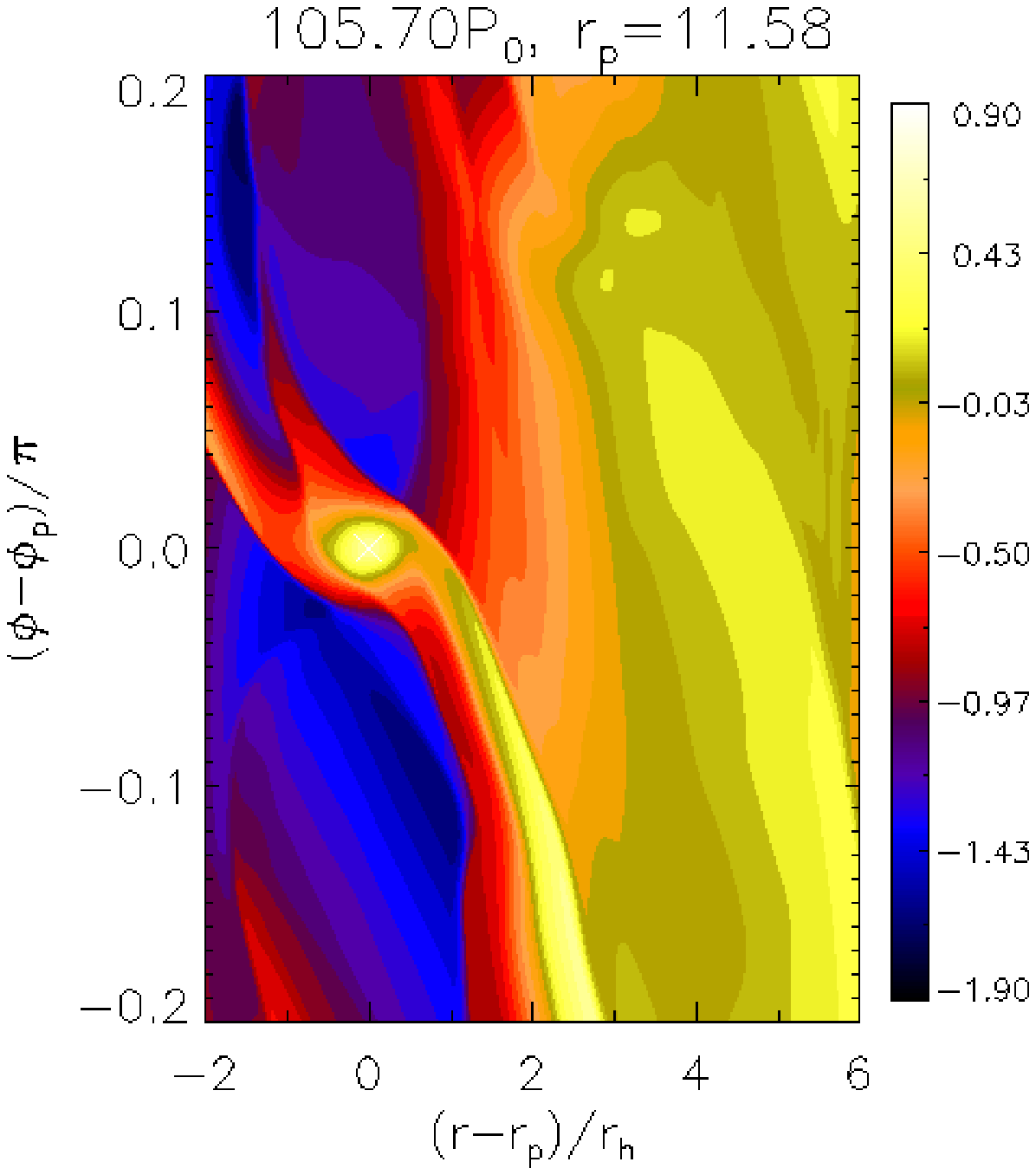}
\caption{Two examples of how the passage of an 
 edge mode spiral arm by the planet can increase its orbital radius. 
The disc model is $Q_o=1.7$ and the logarithmic relative surface density perturbation
   $\log{\left[\Sigma/\Sigma(t=0)\right]}$ is shown.
\label{ch4_Q1d7_polarxy2}}
\end{figure}

\subsection{Surface density asymmetry}

In Fig. \ref{ch4_Q1d7_polarxy2} we plot two interaction events between an
edge mode spiral arm and the planet. 
In the first, we see that the spiral wave disturbance  
extends into within $2r_h$ upstream of the planet. This causes a surface
density asymmetry ahead and behind the planet,  the configuration here corresponding to 
positive coorbital torque. This is more apparent in the second event: over-density
builds up just ahead of the planet as material undergoes inward horseshoe turns. The fluid shocks  
while executing the U-turn, since giant planets induce shocks close to their orbital radii
\citep{lin10}. 
A stable gap would have been cleared of material by a Jovian mass planet and such an 
over-density ahead of the planet would not exist. 

\begin{figure}
\centering
\includegraphics[scale=.24,clip=true,trim=0cm 0cm 1.83cm
0cm]{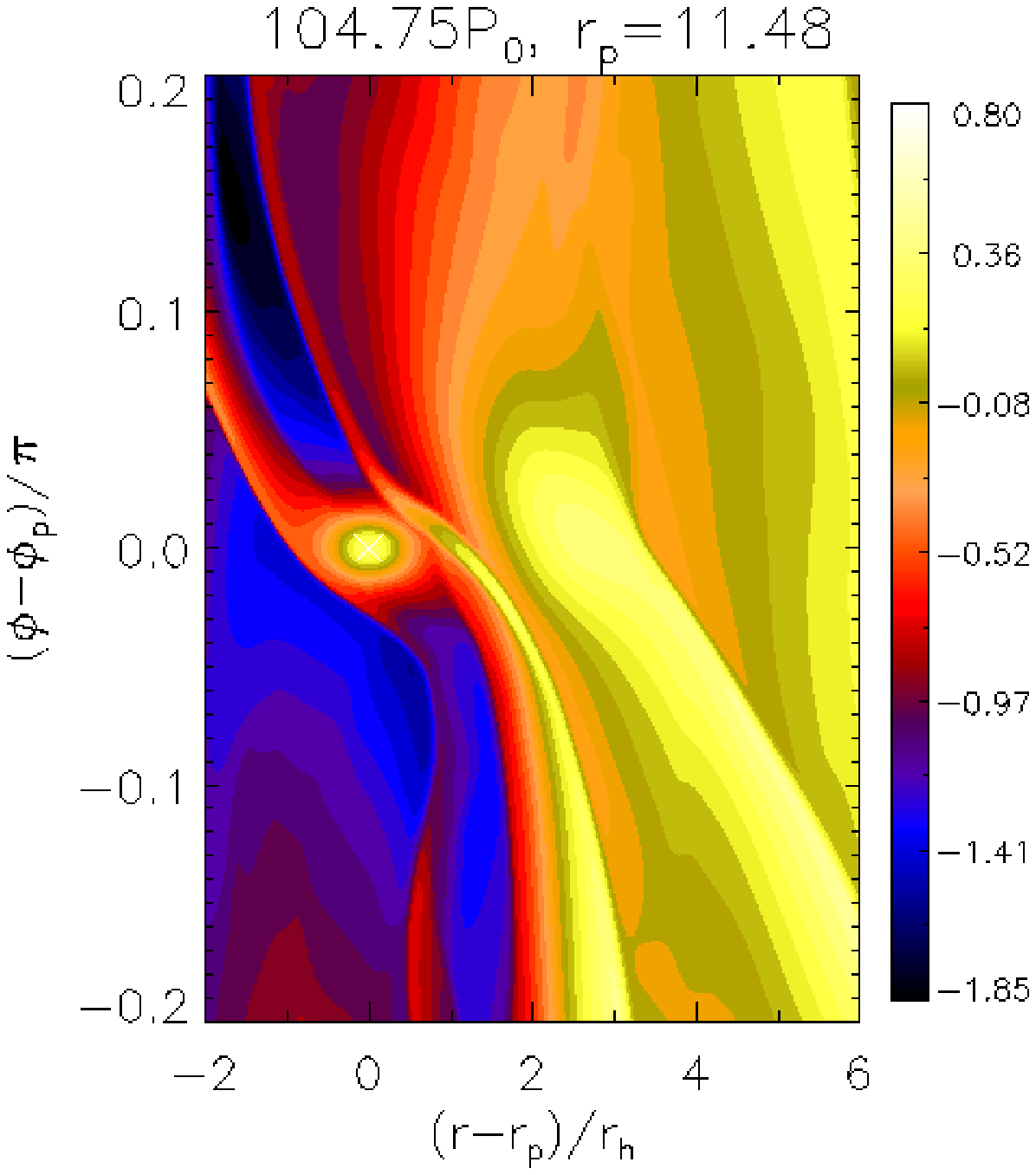}\includegraphics[scale=.24,clip=true,trim=2.23cm
0cm 1.83cm
0cm]{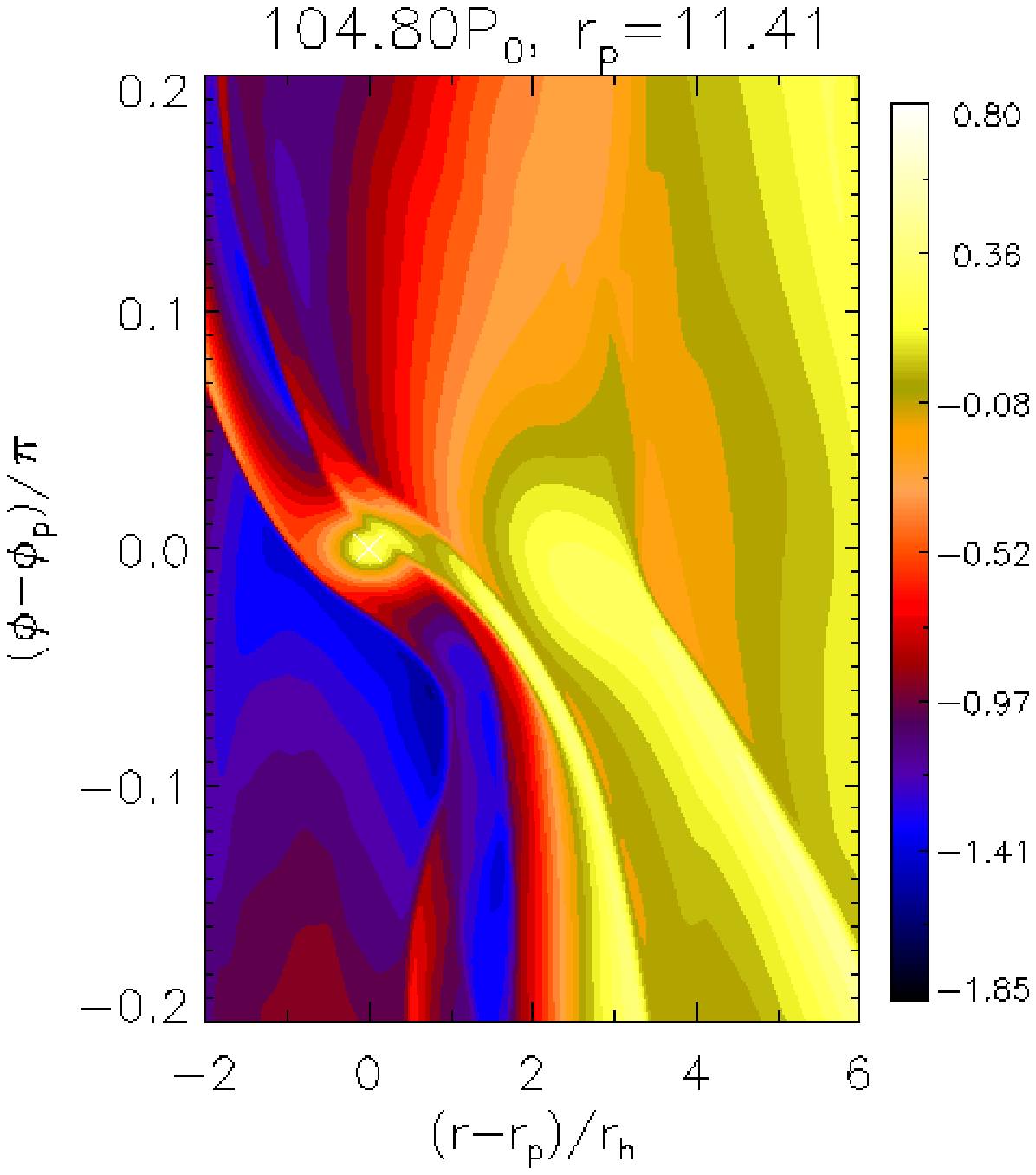}\includegraphics[scale=.24,clip=true,trim=2.23cm
0cm 1.83cm 0cm]{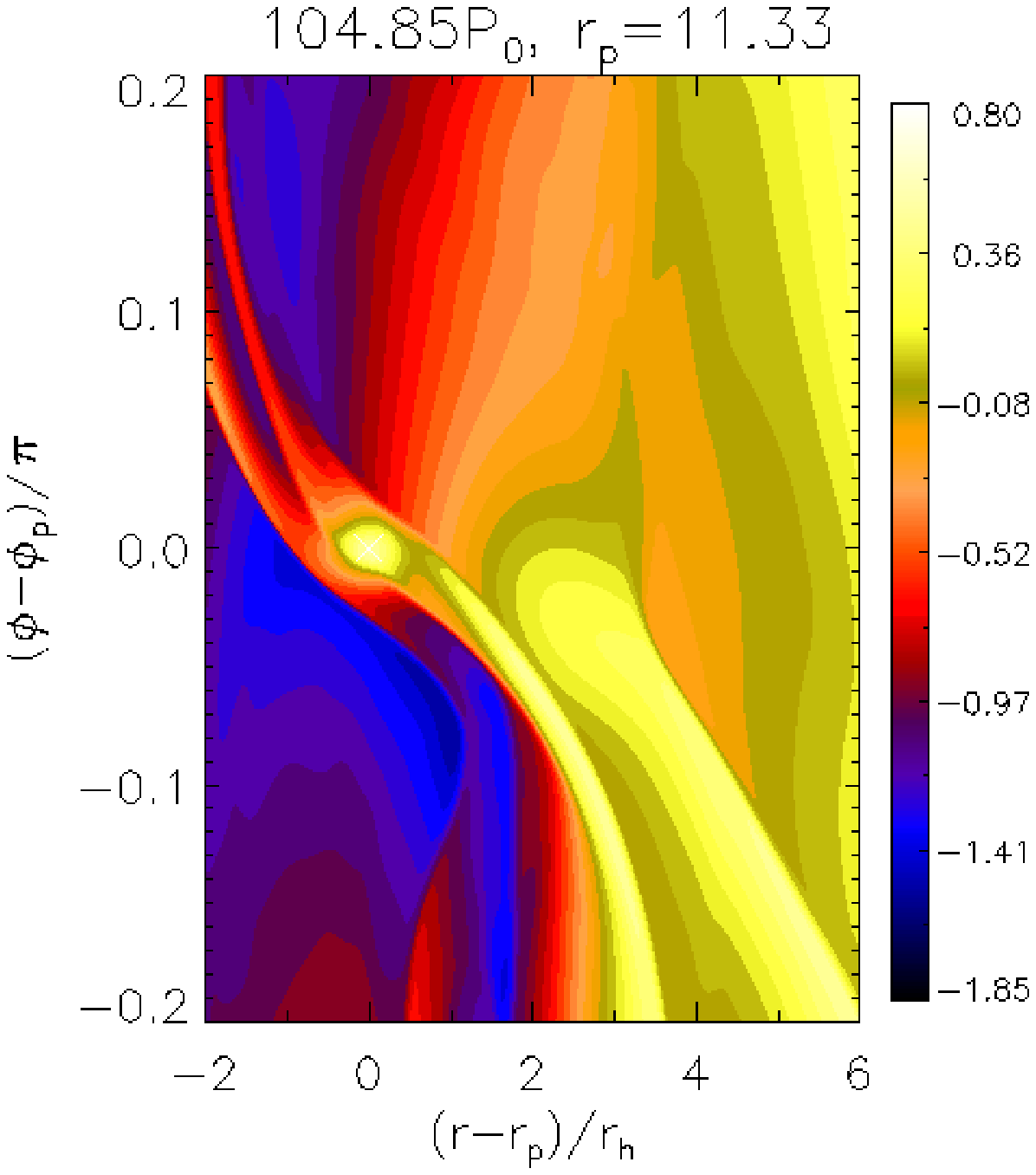}\includegraphics[scale=.24,clip=true,trim=2.23cm
0cm 0cm 0cm]{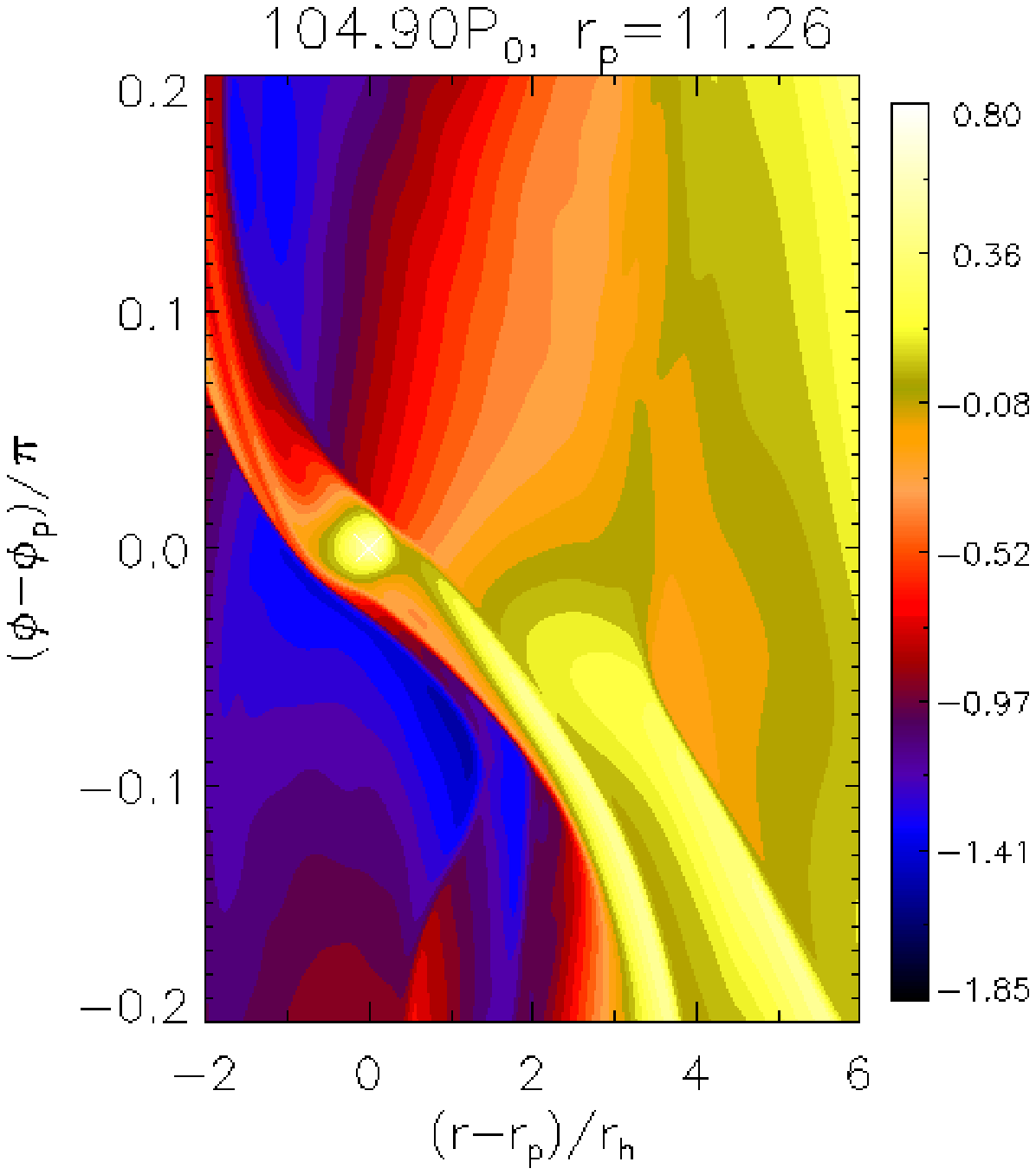}
\caption{Passage of an edge mode spiral arm by the planet in the $Q_o=1.7$ disc model. 
The logarithmic relative surface density perturbation
   $\log{\left[\Sigma/\Sigma(t=0)\right]}$ is shown. In this event the orbital radius decreases 
 because the planetary wake is effectively enhanced by the edge mode. Nevertheless, like
Fig. \ref{ch4_Q1d7_polarxy2}, this plot 
also demonstrates how edge modes can contribute positively to the torque
by supplying material into the coorbital region
for the planet to scatter inwards. 
\label{ch4_Q1d7_polarxy3}}
\end{figure}

It is important to note that we have chosen the snapshots in Fig. \ref{ch4_Q1d7_polarxy2}, where 
the orbital radius $r_p$ increases, to demonstrate how edge modes
can provide a positive torque. However, not every passage of the spiral arm 
increases $r_p$, as implied by the non-monotonic migration. An example is shown in 
Fig. \ref{ch4_Q1d7_polarxy3}. Since the outer planetary wake
is associated with negative Lindblad torques, surface density perturbations due to 
edge modes may enhance it when the two overlap. 

The overall outward migration implies positive torques produced by edge modes are on average  
more significant. This may not be surprising since the positive torque comes from material
crossing $r_p$. This is also the mechanism for type III migration \citep{masset03} which can be 
much faster than migration due to Lindblad torques.


\subsection{Torques and mode amplitudes}
\begin{figure}
\centering
\includegraphics[scale=0.42,clip=true,trim=0cm 1.5cm 0cm 0cm]{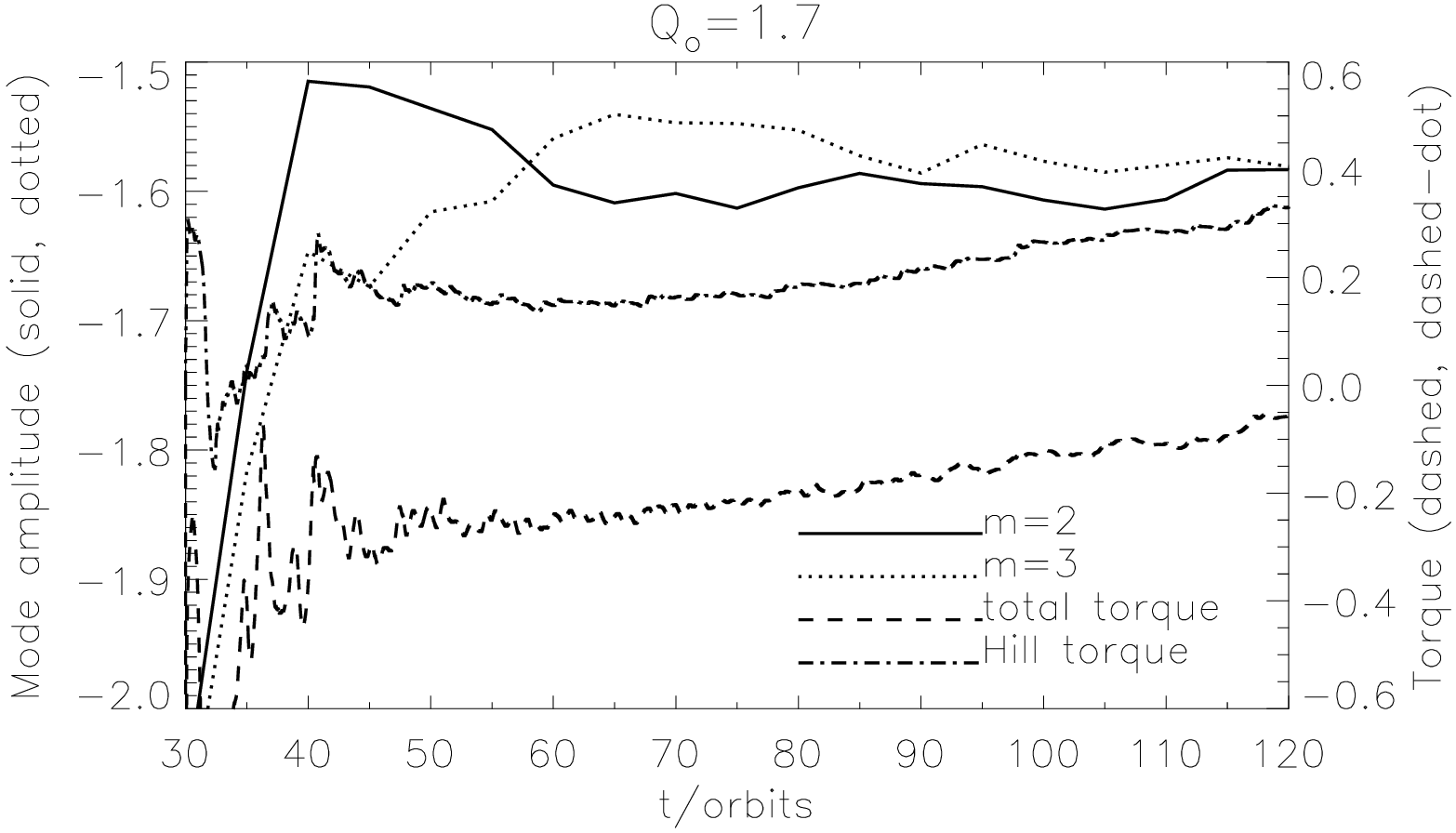}\\
\includegraphics[scale=0.42]{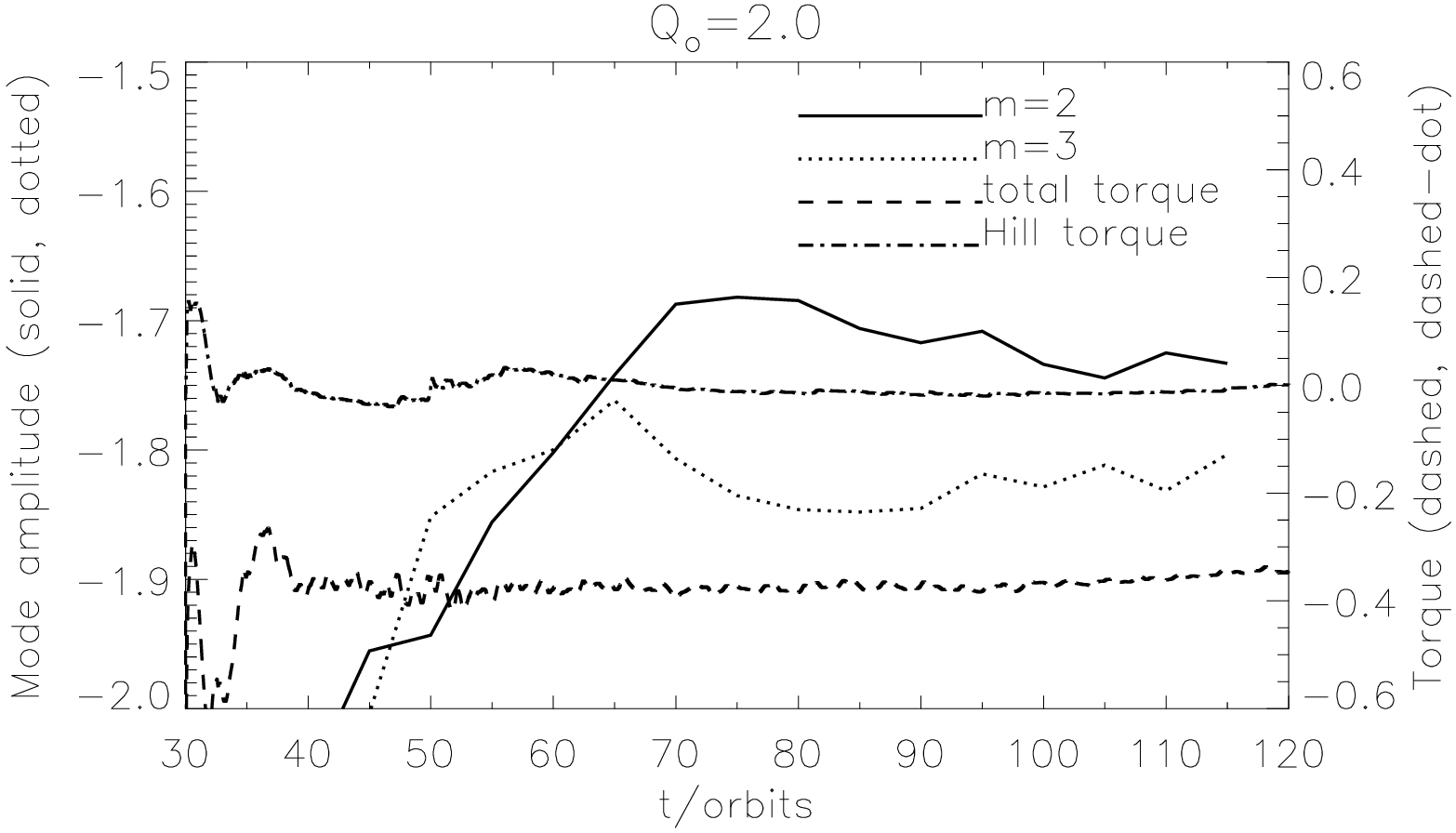}
\caption{Evolution of the amplitude of the $m=2$ (solid) and $m=3$
  (dotted) modes in surface density in the outer disc
  $r\in[r_p,r_o]$ in comparison to the total disc-on-planet torque
  (dashed) and the torque contribution from the Hill sphere
  (dashed-dot). All quantities are running-time averages. Fourier modes
  have been normalised by the axisymmetric amplitude and are plotted on
  a logarithmic scale. The upper
  plot is $Q_o=1.7$, in which the planet migrates outwards. The lower
  plot is the non-migrating $Q_o=2.0$ case.
\label{ch4_Q1d7_mode_torque}}
\end{figure}

To emphasise the significance of edge modes on disc-planet torques we can
compare the fiducial case $Q_o=1.7$ to the non-migrating 
case $Q_o=2.0$. Fig. \ref{ch4_Q1d7_mode_torque} shows the time-averaged evolution of
non-axisymmetric modes in the outer disc ($r>r_p$) and torques in these two cases. 
The amplitude of the $m^\mathrm{th}$ Fourier mode is defined as 
$|C_m/C_0|$,  where
\begin{align}
C_m = \int_{r_p}^{r_o}dr\int_0^{2\pi}d\phi \Sigma(r,\phi)\exp{(-\mathrm{i}m\phi)}, 
\end{align}
and we plot $|C_m/C_0|$ for $m=2,\,3$.

The modes have larger amplitudes for $Q_o=1.7$ and the total torque
increases with time \footnote{The total torque values
are negative, despite overall outward migration, because the running
time-average is plotted. This average includes the phase of planet release, when the disc-planet
torque is large and negative.}. The figure shows
that this is attributed to torques from within the
Hill radius. 
 By contrast, although the $Q_o=2.0$ case also develops large-scale spirals later on, they are of
smaller amplitude and do not make the Hill torque positive. The
total time-averaged torque remains fairly constant. 

Fig. \ref{ch4_Q1d7_mode_torque}, along with Fig. \ref{ch4_Q1d7_polarxy}---\ref{ch4_Q1d7_polarxy3}, confirms that large-scale
spirals which extend into the gap can provide significant torques. Edge modes naturally fit this 
description since they are associated with vortensity maxima just inside the gap. 
Because the outer disc is  
more unstable, the corotation radius $r_c$ of edge modes lies beyond $r_p$ so
its pattern speed is smaller than the planet's rotation. 
Thus, over-densities associated with corotation approach the planet from upstream, 
but because $r_c$ is actually within the planet's coorbital region, the associated
fluid elements are expected to execute inward horseshoe turns upon 
approaching the planet. This provides a positive torque on the planet, and if
the edge mode amplitude is large enough, it may reverse the usual
tendency for inward migration. 

The picture outlined above is consistent with the fact that edge modes
are have corotation at vortensity maxima. A giant planet can induce 
shocks very close to its orbital radius  and  
vortensity is generated as fluid particles execute horseshoe turns across
the shock. That is, vortensity maxima are associated with fluid  
on horseshoe orbits. So when non-axisymmetric disturbances associated 
with vortensity rings develop (i.e. edge modes), the associated over-densities can be 
expected to execute horseshoe turns. 



\subsection{Torque distribution}
\begin{figure}
    \centering
    \includegraphics[scale=0.42,clip=true,trim=0cm 1.75cm 0cm
    0cm]{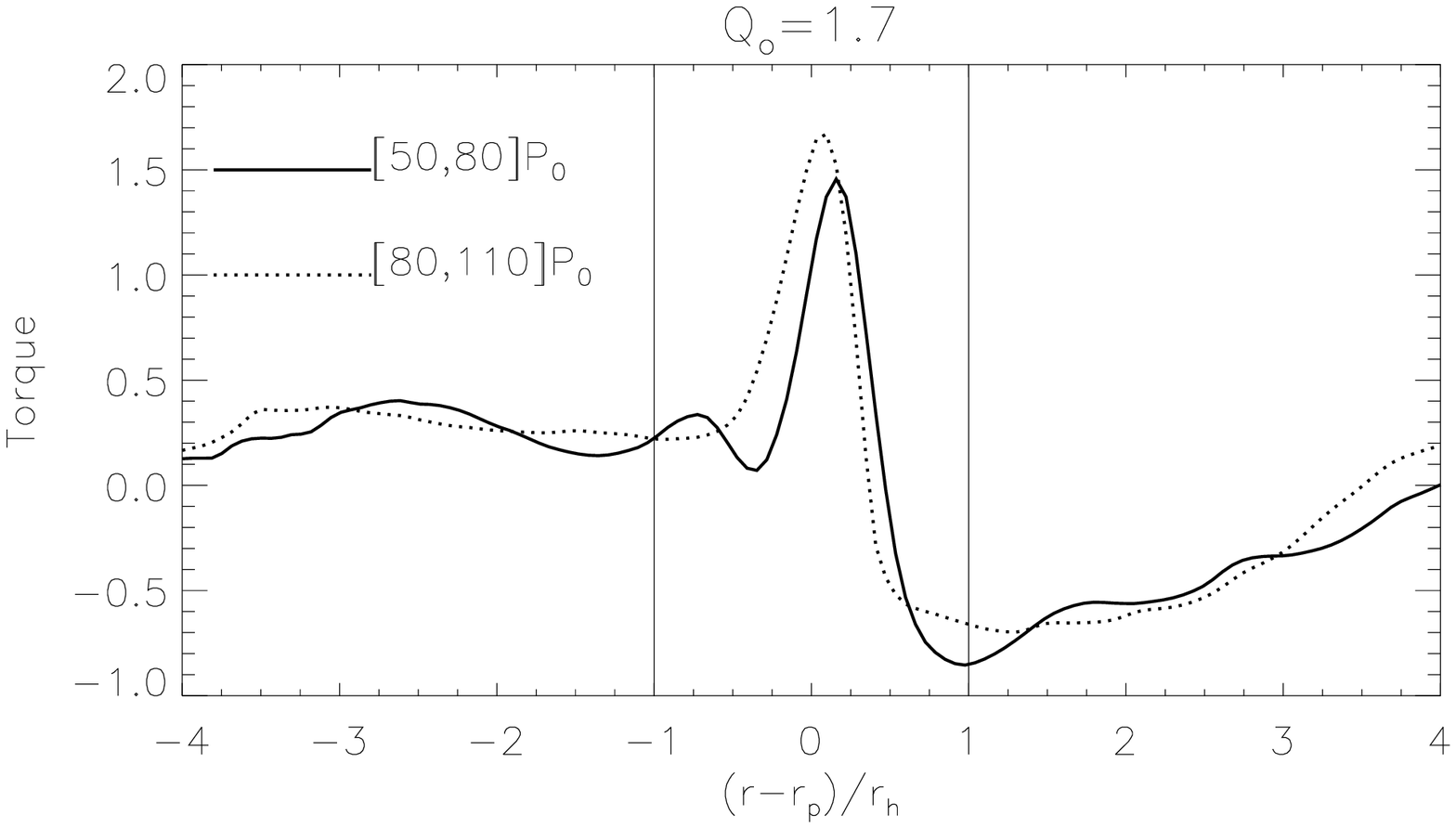}\\
    \includegraphics[scale=0.42]{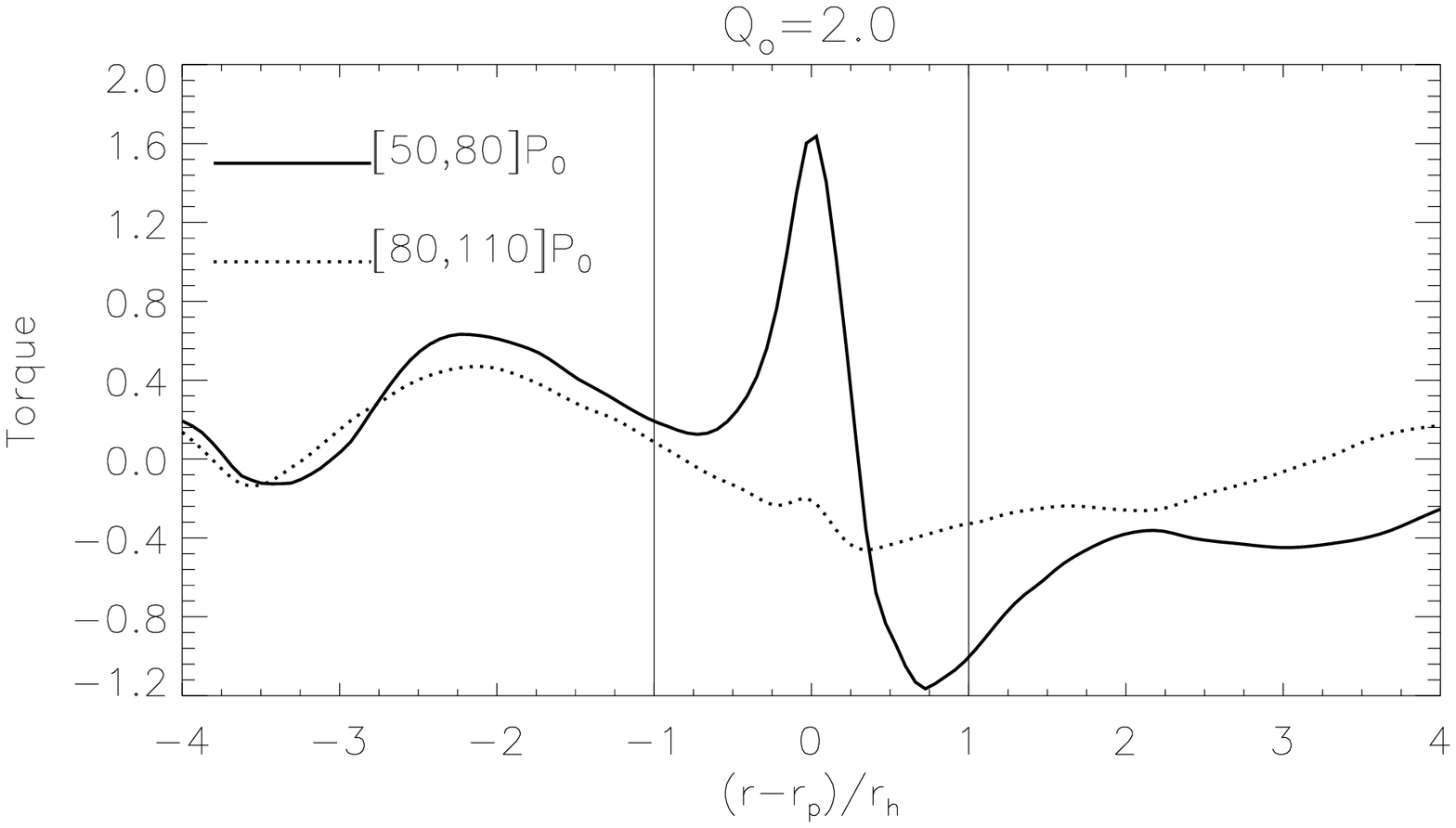}
    \caption{Torque densities averaged over $t\in[50,80]P_0$ (solid) and
      over $t\in[80,110]P_0$ (dotted). The upper plots are from the $Q_o=1.7$ case,
      which display outward migration and the lower plots are from the
      non-migrating $Q_o=2.0$ case. Vertical lines in each plot indicate
      the Hill radius.
      \label{ch4_torque_contrib}}
  \end{figure}

The analysis above suggests that torques from within the gap are responsible for
the gradual increase in $r_p$ or $a$. 
Fig. \ref{ch4_torque_contrib} compares torque densities in
the fiducial case to the non-migrating case. The torques have been
averaged over $30P_0$ at two time intervals. 

For $t\in[50,80]P_o$
radial plots for both cases show a positive torque at
$r_p$. By $t\in[80,110]P_0$ this torque has
diminished for $Q_o=2.0$. This is expected for Jovian planets as
they open a clean gap leaving little material near $r_p$ to torque the planet. 
However, for $Q_o=1.7$, this torque is maintained 
(and even slightly increased) by the edge modes because they bring 
material into the co-orbital region.  Note that the planet in the $Q_o=1.7$ case 
is migrating outwards for $t\in[80,110]P_0$, so migration itself may 
also contribute to sustaining the torque. 


This torque is caused by over-density ahead in comparison to that 
behind the planet, i.e. material flowing radially inwards across 
the planet's orbital radius, as seen in Fig. \ref{ch4_Q1d7_polarxy2}.  
However, unlike the single scattering events by vortices or spirals described in
\cite{lin10,lin11b}, which dominate most of the migration, here one
has to average over many spiral-planet interaction events to see the 
migration.

\section{Summary and discussion}\label{concl}
We have performed self-gravitating disc-planet 
simulations to see the effect of large-scale spiral modes associated 
with a gap opened by a giant planet on its migration. 
In our disc models, edge modes lead to outward migration over timescales of a 
few tens of initial orbital periods. This contrasts to standard type II 
inwards migration on viscous timescales \citep{lin86}. 
This difference demonstrates 
that gap instabilities significantly affect planetary migration in massive discs. 

It is important to note that we have specifically considered 
unstable planetary gaps. This 
differs from disc models employed by \cite{baruteau11} 
and \cite{michael11}, which are at least twice as massive as our fiducial case and 
develop gravitational instabilities without a planet. Furthermore, 
the planet does not open a gap in their models. By contrast, gravitational activity
in our discs are entirely due to the planet opening a gap. 
The instability then back-reacts on planet migration.

\cite{baruteau11} argued that a single giant planet in a massive, 
gravito-turbulent disc effectively undergoes type I migration, which resulted
in rapid inwards migration in their models. This again contrasts to our less massive 
discs, which are not gravito-turbulent and the interaction we consider 
is that between a planet and large-scale spiral waves associated with gap edges.

In the planet's frame, an edge 
mode spiral, with corotation at a vortensity maximum exterior to $r_p$, 
approaches the planet from $\phi > \phi_p$. 
However, the disturbance extends into the planet's 
coorbital region, so there are associated over-densities 
that execute inward horseshoe turns, which
provide a positive torque. The resulting outward migration means that
this positive torque must on average exceed any negative torques. 

In order to sustain this positive torque, edge modes need to be sustained 
to supply material to the planet for interaction. 
This in turn requires the  existence of vortensity maxima,
despite the development of edge modes destroying them. 
However, they are easily regenerated by giant planets because they 
induce shocks which act as a source of vortensity \citep{lin10}. 

Planetary migration observed here closely resembles classic type
III migration because it relies on 
torques from the coorbital region and the accelerated increase in
semi-major axis indicates a positive feedback
\citep{masset03,peplinski08c}.  Accelerated migration may be
  expected because edge modes move the planet toward regions
  of higher surface density (the gap edges) 
so more material enters the planet's co-orbital region. However, other
outcomes are also conceivable (see \S\ref{issues}). 

The main difference from the type III migration originally described by 
\citeauthor{masset03} is 
that the flow across the planet's orbit is non-smooth, because edge
modes provide distinct fluid blobs, rather than a 
continuous flow across $r_p$. Our results above may be
  interpreted as a discontinuous runaway type III migration.

Also, where type III migration usually applies to
partial gaps and therefore Saturn-mass planets, the Jovian mass planet
used here resides in a much deeper gap. 
Radial mass flux across the gap is still possible, despite the
gap-opening effect of Jovian planets, because the unstable edge modes
protrude the gap edge. 

Our understanding of the effect of edge modes
is based on the geometry of the gap structure. That is, the existence of a
vortensity maximum (equivalently $Q$-maximum) close to the gap edge, and its 
association with global spiral modes. Thus the mechanism we have identified, that
edge modes bring material to the planet for interaction, should be a robust
phenomenon for planetary gaps in massive discs.

\subsection{Comparison to previous work}

Migration here differs to scattering by an edge mode spiral arm
described  by \cite{lin11b}, in which a single interaction increases
the orbital radius of a Saturn mass planet by $40\%$ over a timescale
of only $4P_0$, moving it out of its gap (see their Fig. 20).  By
contrast, the 2 Jupiter mass planet simulated here  migrates outwards
by $20\%$ over a few 10's of $P_0$, during most of which it remains
in a gap, and there are multiple encounters with edge mode spiral
arms.  

In both cases the torque comes from material crossing the planet's
orbital radius, but interaction occurs more readily with the less massive Saturn mass
planet since its gap-opening effect is weaker. 
Furthermore, the equation of state employed in the preliminary
simulation of 
\cite{lin11b} allows high surface densities near the planet,
potentially increasing the coorbital torque magnitudes.  

Note that if the planet is first allowed to move after edge modes
develop, then the initial migration direction would depend on the
relative position of the edge mode spiral with respect to the
planet. This was the case in \cite{lin11b}, the planet being released
when a spiral arm is just upstream. 

However, for the models discussed
in this paper, edge modes do not develop before planet release
($t=30P_0$). This can be seen by the well-defined $\mathrm{max}(Q)$
, equivalently vortensity maximum, in the gap profiles in Fig. \ref{ch4_1D_profile}. 
The onset of edge modes would have destroyed this local maximum
\citep[][Fig. 14]{lin11b}.

\subsection{Implications}

A consequence of the above is that formation of a clean gap is expected to
become increasingly difficult in massive discs. Planetary gaps
correspond to existence of vortensity maxima, but
these become unstable to edge modes in massive discs and instability
tends to fill the gap. Therefore the standard picture and formulae for 
type II migration should not be applied to gap 
opening planets in massive protoplanetary discs. 

We remark that migration is not only relevant to
planets in protoplanetary discs. Analogous interactions 
have been discussed in the context of stars in 
black hole accretion discs \citep{kocsis11, mckernan11}. Self-gravity could be important
in outer regions of these discs, so our results may also be relevant 
to these situations.

\subsection{Outstanding issues}\label{issues}
Our focus in this paper was to identify the 
mechanism by which edge modes may torque up the planet. 
We have thus considered one particular disc model and only 
varied one parameter, the surface density scale. Although the 
simulation timescales were sufficient for our purpose,
these are short compared to disc lifetimes.

We expect inter-dependencies between planet, disc properties 
and migration for the specific situation of a planet interacting
with a gravitationally unstable gap, which was induced by it 
in the first place. In our fiducial 
simulation the planet eventually goes into classic type III
migration.  However, if the planet moves out and leaves its gap
  with a slow enough migration speed, it may open a new gap which develops edge
 modes\footnote{Edge modes also require the existence of a wider disc
   beyond the gap edge, so outward
migration should become ineffective once the planet moves close the
disc outer boundary.} 
and induces further outward migration, or even fragmentation
(\S\ref{model_stability}).
It is also conceivable for some disc parameters, the positive torque from edge modes
 are sufficiently small, so that it may be balanced by inwards type II
 migration on 
longer timescales.  
Possible long term outcomes need to be
explored in a parameter study.

It should be noted that our adopted disc model is biased
towards outward migration because the Toomre $Q$ decreases outwards
(approximately like $r^{-1/2}$) so edge modes associated with the
outer disc develop preferentially. 
This is consistent with the expectation
that typical discs become more self-gravitating with increasing radius.   
However, if the disc model was such that the inner disc is more unstable, 
then edge modes should promote inward migration. 
And if independent edge modes of comparable amplitude develop on either side 
of the gap there could be little migration overall.

Significant torques originate from material close to the
planet, so numerical treatment of the Hill sphere is a potential
issue. This concern also applies to standard type III migration, but  
we remark that the adopted equation of state was originally designed for numerical 
studies of type III migration \citep{peplinski08a}. Furthermore, 
the fully self-gravitating discs considered here are what is required
for accurate simulations of type III migration \citep{crida09}. 
As a check we have also performed lower resolution simulations 
which also resulted in outward migration. 
 

This EOS mimics heating near the planet but is not a quantitative model for 
the true thermodynamics. If this EOS under-estimates the heating, then the positive 
torque could be over-estimated and vice versa. On the other hand, edge mode corotation
resides outside the Hill radius but still inside
the gap. Thus the numerical treatment within the Hill radius does not
affect existence of edge modes and their associated over-densities
should affect torques originating from the coorbital region 
in the way described in this paper.   




\end{document}